\documentclass[10pt,a4paper,pre,showpacs,superscriptaddress,bibnotes]{revtex4}
\usepackage{bm}
\usepackage{hyperref}
\usepackage{graphicx}
\usepackage{amsmath}
\usepackage{amssymb}
\usepackage{stmaryrd}
\usepackage{color}

\newcommand{\ra}{\rightarrow}

\newcommand{\yaming}[1]{\textcolor{black}{#1}}

\begin{document}


\title{Weak-noise limit of a piecewise-smooth stochastic differential equation}
\author{Yaming Chen}
\email{yaming.chen@qmul.ac.uk}
\affiliation{School of Mathematical Sciences, Queen Mary University of London, London E1 4NS, United Kingdom}

\author{Adrian Baule}
\email{a.baule@qmul.ac.uk}
\affiliation{School of Mathematical Sciences, Queen Mary University of London, London E1 4NS, United Kingdom}

\author{Hugo Touchette}
\email{htouchet@alum.mit.edu}
\affiliation{National Institute for Theoretical Physics, Stellenbosch 7600, South Africa}
\affiliation{Institute of Theoretical Physics, University of Stellenbosch, Stellenbosch 7600, South Africa}

\author{Wolfram Just}
\email{w.just@qmul.ac.uk}
\affiliation{School of Mathematical Sciences, Queen Mary University of London, London E1 4NS, United Kingdom}

\date{October 11, 2013}

\begin{abstract}
We investigate the validity and accuracy of weak-noise (saddle-point or instanton) approximations for piecewise-smooth stochastic differential equations (SDEs), taking as an illustrative example a piecewise-constant SDE, which serves as a simple model of Brownian motion with solid friction. For this model, we show that the weak-noise approximation of the path integral correctly reproduces the known propagator of the SDE at lowest order in the noise power, as well as the main features of the exact propagator with higher-order corrections, provided the singularity of the path integral associated with the nonsmooth SDE is treated with some heuristics. We also show that, as in the case of smooth SDEs, the deterministic paths of the noiseless system correctly describe the behavior of the nonsmooth SDE in the low-noise limit. Finally, we consider a smooth regularization of the piecewise-constant SDE and study to what extent this regularization can rectify some of the problems encountered when dealing with discontinuous drifts and singularities in SDEs.
\end{abstract}

\pacs{02.50.--r, 05.40.--a, 46.55.+d, 46.65.+g}
\maketitle

\section{Introduction}
\label{sect.introduction}

The study of piecewise-smooth differential equations is a relatively recent topic in field dynamical systems \cite{filippov1988,bernardo2008,makarenkov2012,biemond2012,colombo2012}, despite their wide use in applied sciences and engineering. The dynamics generated by these equations displays many unexpected phenomena, including stick-slip transitions associated, for example, with solid friction forces \cite{bowden1950,elmer1997,persson1998,berger2002} and bifurcations that do not appear in the standard classification of catastrophes of smooth dynamical systems \cite{makarenkov2012,biemond2012,colombo2012}. They also show, in the case of systems with discontinuous derivatives or forces (so-called Filippov systems \cite{filippov1988}), multivalued solutions for a given initial condition, leading to a loss of determinism \cite{jeffrey2011}.

Stochastic versions of piecewise-smooth dynamical systems perturbed by noises are also used as models of physical and biological systems. Stochastic differential equations (SDEs) with piecewise-smooth drifts are commonly used, for example, in stochastic or Brownian ratchets, which serve as models of diffusion and transport in a variety of biological motors \cite{reimann2002}. Another important class of problem concerns diffusion of solid objects on solid surfaces, which can be modeled phenomenologically using piecewise-smooth SDEs with solid friction forces \cite{caughey1961,atkinson1968,gennes2005,hayakawa2005}. The dynamics in this case shows stick-slip transitions, as in the noiseless case, but also new features due to the noise such as directed motion in the absence of a mean force bias \cite{baule2012} and noise-dependent decay of correlation functions \cite{touchette2010c}. Some of these features have been investigated experimentally in \cite{chaudhury2008,goohpattader2009,goohpattader2010,gnoli2013}.

The theory of piecewise-smooth SDEs is only in its infancy compared to its noiseless counterpart. Exact solutions for the transition probability distribution or propagator of a few simple piecewise-constant or piecewise-linear SDEs are known (see, e.g., \cite{caughey1961,touchette2010c,
simpson2012,karatzas1984,zhang1990,benes1980,browne1995}). 
Some studies have also looked at the large deviations of SDEs with discontinuous 
drift (so-called SDEs with discontinuous statistics) 
\cite{korostelev1993,chiang2000,boue2000,gradinaru2001}. However,
from these disconnected studies it is not clear how nonsmooth SDEs can be studied with techniques developed 
and used for smooth systems. Here we focus on two such techniques, namely, the path integral representation of propagators and the weak-noise (also called saddle-node or instanton) approximation of these integrals \cite{chaichian2001,graham1989,touchette2009}. The weak-noise limit is particularly interesting from a physical point of view because a piecewise-smooth system does not necessarily behave continuously with the magnitude of a force or noise and may therefore behave in a non-trivial way in the limit of vanishing noise. This raises the question of the validity of the weak-noise limit for nonsmooth systems.

In this paper, we address this question in a practical way by studying a simple piecewise-constant SDE, defined by
\begin{equation}
\dot v(t)=-\mu \sigma(v(t)) +\sqrt{D}\, \xi(t),
\label{EqLangevinOnlyDryFriction}
\end{equation}
where $\sigma(v)$ denotes the sign of the state variable $v(t)$, $\mu$ is a positive constant, and $D$ is the strength of the Gaussian white noise $\xi(t)$
characterized by
\begin{equation}\label{ba}
\langle\xi(t)\rangle=0,\quad \langle\xi(t)\xi(t')\rangle=2\delta(t-t').
\end{equation}
The notation $\langle \cdots \rangle $ stands for the average over the noise. Physically, the SDE (\ref{EqLangevinOnlyDryFriction}) is the simplest model of stick-slip motion where $v(t)$ represents the velocity of a solid object of unit mass sliding over a surface with solid (also called dry or Coulomb) friction coefficient $\mu$ per unit mass \cite{caughey1961,atkinson1968,gennes2005,hayakawa2005}. Mathematically, it is also the simplest piecewise-smooth SDE whose time-dependent propagator $p(v,t|v_0,0)$, representing the probability density that $v(t)=v$ given the initial condition $v(0)=v_0$, is known exactly \yaming{\cite{touchette2010c}}. It is thus a good starting point
for benchmarking results about nonsmooth SDEs.

In the present case, we have
\begin{equation}
p(v_t,t|v_0,0)=\frac{\mu}{D}\, \hat{p}\left(\frac{\mu}{D}v_t,\frac{\mu^2}{D} t \bigg |\frac{\mu}{D} v_0,0\right),
\label{EqanalyticalSolutionDryOnly}
\end{equation}
where
\begin{equation}
\hat{p}(x,\tau|x',0) = \frac{ e^{-\tau/4} }{ 2\sqrt{\pi\tau} }\,  e^{ -(|x|-|x'|)/2 }\, e^{-(x-x')^2/(4\tau)}
 +\frac{e^{-|x|}}{4}\bigg[1+\mathrm{erf}\left(\frac{ \tau-( |x|+|x'| ) }{2\sqrt{\tau}}\right)\bigg]
\label{EqExactSolutionDryFriction}
\end{equation}
is the propagator in nondimensional units and  $\mathrm{erf}(x)$ is the error function. This propagator is obtained by explicitly solving the corresponding time-dependent Fokker-Planck equation, which in nondimensional units has the form
\begin{equation}\label{fpe}
\frac{\partial }{\partial \tau}\hat{p}(x,\tau|x',0)=\frac{\partial }{\partial x}[\sigma(x)\hat{p}(x,\tau|x',0)]+\frac{\partial^2 }{\partial x^2}\hat{p}(x,\tau|x',0).
\end{equation} 
Taking the limit $t\rightarrow\infty$ of $p(v,t|v_0,0)$, we obtain the stationary probability density function of Eq.~(\ref{EqLangevinOnlyDryFriction}), 
\begin{equation}\label{EqStationarySolutionDryFriction}
p(v)=\frac{\mu}{2D}\, e^{-\mu |v|/D},
\end{equation}
which solves the time-independent Fokker-Planck equation \cite{hayakawa2005}. Note that $p(v_t,t|v_0,0)$ and, by extension, $p(v)$ are symmetric under the
change $ v_0\rightarrow -v_0 $ and $ v(t)\rightarrow -v(t) $ due to the symmetric force $ \sigma(v) $.
Thus we can confine the analysis to the case $v_0>0$ without loss of generality.

In the next section, we compare these exact results, which are valid for any noise power $D$, with various 
low-noise approximations of the path integral representation of $p(v,t|v_0,0)$ 
in order to test the validity and accuracy of these approximations and to discuss subtle singularities arising in the path integral when dealing with discontinuous drifts. 
Related path integrals were studied in \cite{baule2010, baule2011} for a non-exactly solvable model
of solid friction.
For the model that we consider, we will see that the weak-noise approximation gives the correct propagator at the lowest order in $D$, as well as its main features with higher-order corrections, provided some heuristics are used to treat the singularities of the path integral. The higher-order corrections are able, in particular, to reproduce the short- and long-time behavior of the propagator, as well as its tail behavior.

To treat the singularity of the system in a more explicit and systematic way, we then consider in Sec.~\ref{sect.regularization} a regularized version of our model, given by
\begin{equation}
\dot{v}(t)=-\mu \tanh\left(\frac{v(t)}{\varepsilon}\right)+\sqrt{D}\, \xi(t),
\label{EqLangevinTanh}
\end{equation}
which recovers the original piecewise-smooth model of Eq.~(\ref{EqLangevinOnlyDryFriction}) in the limit $\varepsilon\rightarrow 0$. Although we do not have an explicit expression for the propagator of this smooth (but nonlinear) model, we show with Langevin simulations that the weak-noise approximation of the path integral, which is now well-defined and shows no singularity, reproduces the main features of the propagator at different orders of approximation, with roughly the same accuracy as for the singular model. The regularized model also allows us to obtain analytical results about the instanton or optimal path of the system, which is the most probable path singled out by the weak-noise approximation, and the so-called action functional or quasipotential, obtained by approximating the propagator at the lowest order in $D$ with the optimal path \cite{graham1989,touchette2009,freidlin1984}. From these results, presented in Sec. \ref{sect.analyticalAction}, we are able to study the evolution of the propagator's tails in the long-time limit and to understand the behavior of the SDE in the noiseless limit.  Conclusions drawn from these results are given in Sec.~\ref{sect.conclusion}.

\section{Piecewise-smooth model}
\label{sec.dryfriction}

In this section, we compare the exact propagator of the piecewise-constant SDE of Eq.~(\ref{EqLangevinOnlyDryFriction}) with various approximations of the path integral representation of this propagator so as to discuss the validity of these approximations for a nonsmooth SDE. The path integral has the form
\begin{equation}
\label{EqDryOnlyPathIntegral}
p(v_t,t|v_0,0)
=\int_{(v_0,0)}^{(v_t,t)} \mathcal{D}[v]\, J[v]\, e^{-S^{(0)}[v]/( 4D )}
\end{equation}
and involves two terms: the action functional
\begin{equation}
S^{(0)}[v]=\int_0^t [\dot v(s)+\mu \sigma(v(s)) ]^2\,\mathrm{d}s,
\label{EqDryonlyZeroOrderAction}
\end{equation}
which is a measure of the probability of a path $\{v(s)\}_{s=0}^t$ in velocity space, and the Jacobian functional $J[v]$, which is the Jacobian of the transformation of the Gaussian white noise $\xi(t)$ to the $v(t)$ process (see Appendix~\ref{App.pathIntegral}).

For nonsmooth SDEs, two problems arise with the path integral (\ref{EqDryOnlyPathIntegral}). The first is that, since the noiseless system $\dot v=-\mu \sigma(v)$ admits in general piecewise-linear trajectories that are continuous but nondifferentiable at points where $v(t)$ vanishes, the minimization of the action must also be carried over these trajectories, which means that care must be taken with the $\dot v$ term in $S^{(0)}[v]$. The second problem is that the Jacobian, as given in Eq.~(\ref{eqappjac1}) of Appendix~\ref{App.pathIntegral}, is singular. For our model, we formally have  
\begin{equation}
J[v]=\exp\left(\mu\int_0^t\delta(v(s))\, \mathrm{d} s\right) \, ,
\end{equation} 
where $\delta(v)$ is the Dirac delta function. Below we show how to treat this singular contribution and how its inclusion or noninclusion in the saddle-point approximation of the path integral determines different orders of approximation of the propagator as $D\ra 0$. 

\subsection{Zeroth-order saddle-point approximation}
\label{subsecDryonly0}

The lowest order approximation of the propagator $p(v,t|v_0,0)$ is obtained in the noiseless limit $D\ra 0$ by finding the path $\{v^{(0)}(s)\}_{s=0}^t$ that minimizes the action $S^{(0)}[v]$ so as to write
\begin{equation}
p^{(0)}(v,t|v_0,0)=N e^{-S^{(0)}[v^{(0)}]/(4D)},
\label{eqspa0}
\end{equation} 
where $N$ is a normalization constant. We refer to this approximation as the zeroth-order saddle-point approximation [SPA(0)]. The rationale for this approximation is that, in the limit $D\ra 0$, the path integral (\ref{EqDryOnlyPathIntegral}) is dominated by the probability of the most probable or optimal path $\{v^{(0)}(s)\}_{s=0}^t$ having minimal action. The Jacobian $J[v]$ can be neglected at this level of approximation, since it does not depend on $D$. 

The Euler-Lagrange equation associated with the minimization of $S^{(0)}[v]$ in the region $v>0$ and $v<0$ is simply
\begin{equation}
\label{EqEulerLagrangianDryonly}
\ddot{v}^{(0)}=0
\end{equation}
and leads to straight paths
\begin{equation}\label{EqdirectPathDryonly}
v^{(0)}_d(s)=(v_0-v_t)s/t+v_0,
\end{equation}
which we call direct paths. As mentioned above, in addition to these paths, we must consider paths that follow the attractor $v=0$, since these paths appear in the noiseless system. As a result, the minimization of $S^{(0)}[v]$ must be carried out over all continuous and piecewise-linear paths consisting of direct paths and paths following the $v=0$ axis.
This situation differs from smooth SDEs, for which the minimization is generally over all continuously differentiable paths, and leads us to define two important heuristic principles for dealing with nonsmooth SDEs: (i) The action of a path must be evaluated as the sum of the actions of all its linear (or in general smooth) parts without regard to its joining (nonsmooth) points and (ii) any part of a path on the $v=0$ axis (or, in general, on an attractor of the noiseless system) must have a zero action, in analogy with smooth systems.

\begin{figure}
\begin{center}
\includegraphics[scale=0.5]{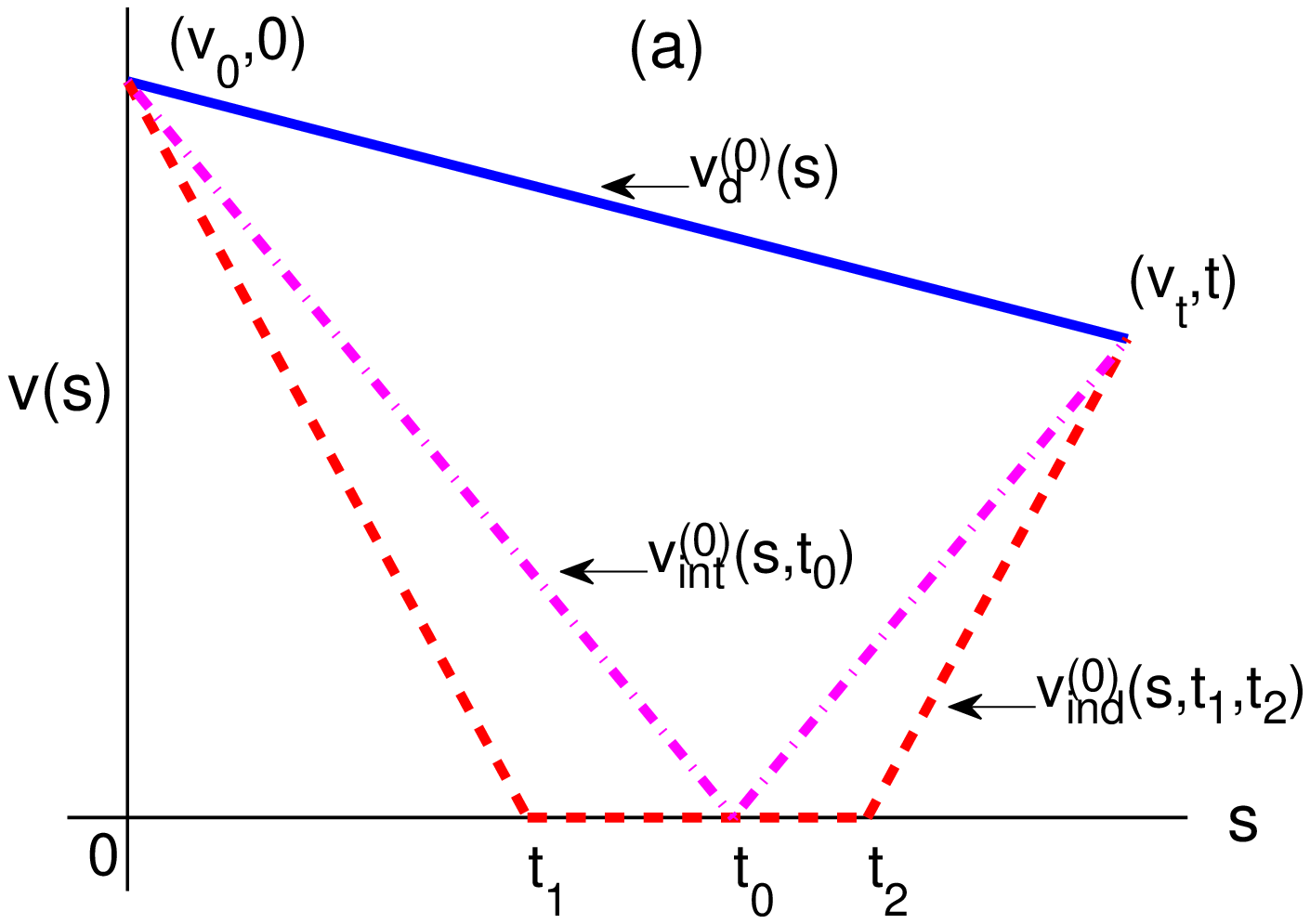}
\includegraphics[scale=0.5]{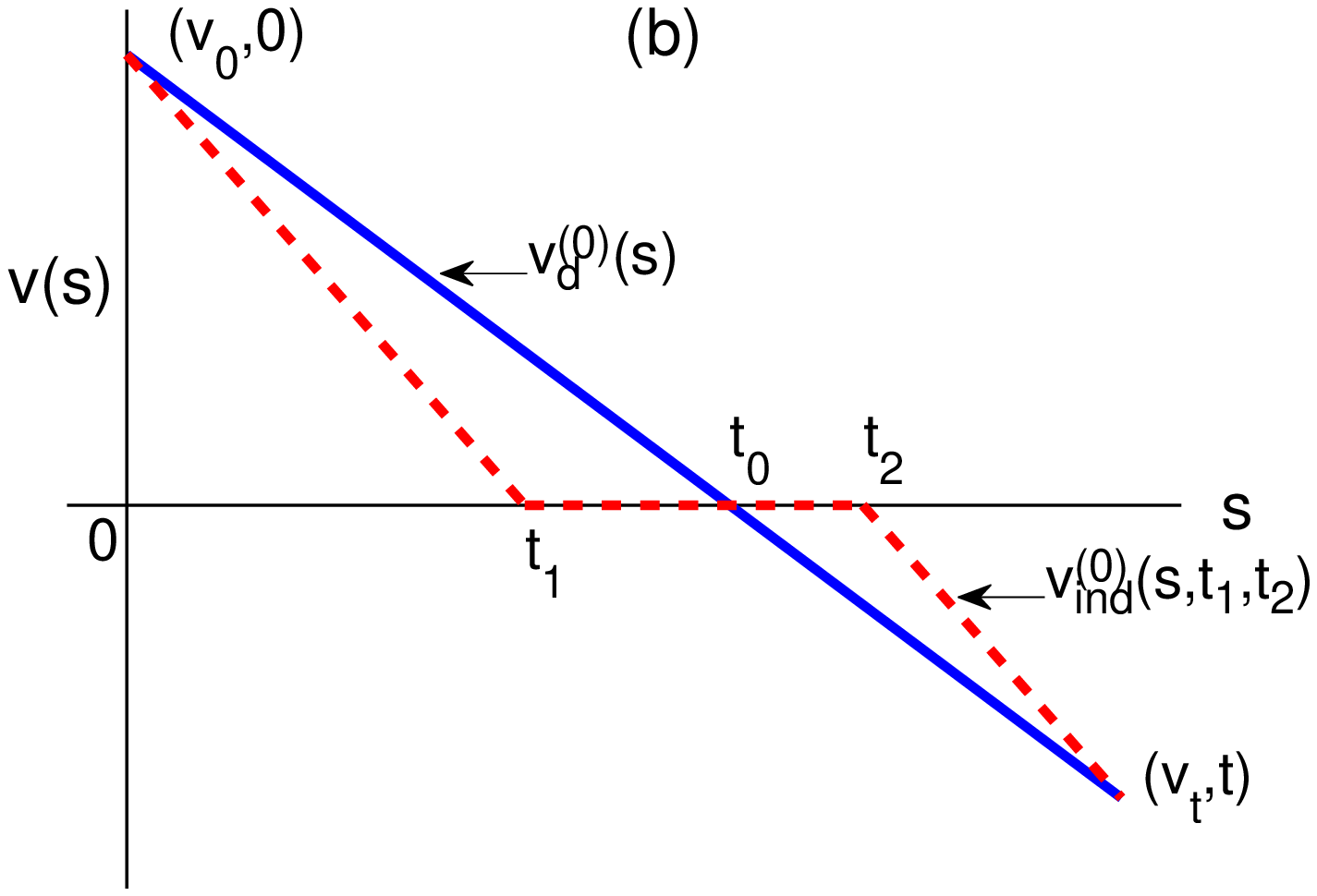}
\end{center}
\caption{(Color online) Different paths considered for minimizing the
action, Eq.~(\ref{EqDryonlyZeroOrderAction}), for (a) $ v_t > 0 $ and (b) $  v_t <0 $.
Here $v_d^{(0)}$ denotes the direct path (\ref{EqdirectPathDryonly}),
$v_{ind}^{(0)}$ the indirect path (\ref{ab}), and 
$v_{int}^{(0)}$ the intermediate path (\ref{ab}) with $t_1=t_2=t_0$.}
\label{FigpathDryonly}
\end{figure}

In the present model, two types of optimal paths arise from the action minimization. The first consists of direct paths $v_d^{(0)}(s)$, found above, which directly link the positive initial velocity $v_0$ to a final velocity $v_t$ [see Fig.~\ref{FigpathDryonly}(a)] and whose action is
\begin{equation}
\label{EqDryonlyDirectZeroOrderAction}
S^{(0)}[v^{(0)}_d]=
\left\{
\begin{array}{lll}
(v_t-v_0+\mu t)^2/t-4\mu v_t & & \mbox{for } v_t < 0\\
(v_t-v_0+\mu t)^2/t & & \mbox{for } v_t> 0.
\end{array}
\right.
\end{equation}
The second type of path is the piecewise linear path mentioned above, consisting of two straight lines in the region $v>0$ or $v<0$ connected by a straight line at $v=0$. The equation of these indirect paths is
\begin{equation}\label{ab}
v_{ind}^{(0)}(s,t_1,t_2)=
\left\{
\begin{array}{lll}
(t_1-s)v_0/t_1 &   & \mbox{for } s < t_1\\
0 &                      & \mbox{for } t_1<s<t_2\\
(s-t_2)v_t/(t-t_2) & & \mbox{for } s > t_2 ,\\
\end{array}
\right.
\end{equation}
where $t_1<t_2$ are arbitrary times at which a path reaches $v=0$ (see Fig.~\ref{FigpathDryonly}). Following the two principles above, we evaluate the action of these paths in a piecewise way with $S^{(0)}[0]=0$ and minimize it for $t_1<t_2$ to obtain
\begin{equation}
\label{EqindirectZeroOrderActionDryonly}
S^{(0)}[v_{ind}^{(0)}]=4\mu |v_t|
\end{equation}
for $ t_1=v_0/\mu$ and $t_2=t-|v_t|/\mu$. Since we require $t_1< t_2$, the lower bound is reached if $|v_t|<\mu t-v_0 $. Note that paths arising in the limit case where $t_1=t_2\equiv t_0$ with $v_0v_t<0$ correspond to direct paths, whereas those that just ``bounce'' on the $v=0$ axis, i.e., $ t_1=t_2\equiv t_0 $ but $v_0v_t>0$, are called intermediate paths [see Fig.~\ref{FigpathDryonly}(a)] and have an action equal to
\begin{eqnarray}\label{DryonlyIntermediateZeroOrderAction}
S^{(0)} [v_{int}^{(0)}]= (|v_t|+v_0-\mu t)^2/t+4\mu|v_t|
\end{eqnarray}
for $t_0=v_0 t/(v_0+|v_t|)$. 

\begin{figure}
\begin{center}
\includegraphics[scale=0.5]{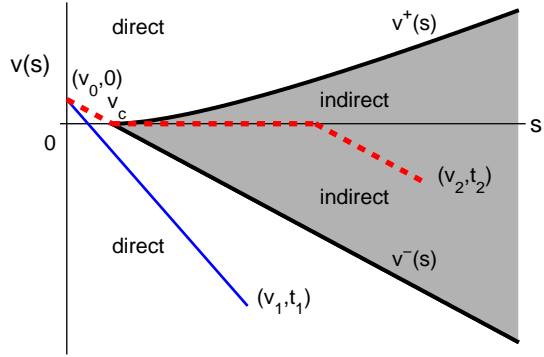}
\end{center}
\caption{(Color online) Regions in the $v_t$-$t$ plane for which the optimal path of the action at leading order (\ref{EqDryonlyZeroOrderAction}) is given by an indirect (direct) path [shaded (white)] [see Eqs.~(\ref{ga})]. Here $ v_c=\mu/v_0 $ denotes the tip of the region.}\label{FigDryonlyPathPlane}
\end{figure}

It can be easily checked that any piecewise-linear paths other than those considered above have a greater action, and so cannot be optimal. Therefore, combining Eqs.~(\ref{EqDryonlyDirectZeroOrderAction}),
(\ref{EqindirectZeroOrderActionDryonly}) and (\ref{DryonlyIntermediateZeroOrderAction})
we can write the equation of the optimal path as
\begin{equation}\label{EqzeroOrderPathDryonly}
v^{(0)}(s)=\left\{
\begin{array}{lll}
v_d^{(0)}(s) & & \mbox{for } \yaming { t < v_0/\mu }\\
v_{ind}^{(0)}(s,v_0/\mu,t-|v_t|/\mu) & & \mbox{for } \yaming { t > v_0/\mu } \mbox{ and } v_t \in [v^-(t),v^+(t)]\\
v_d^{(0)}(s) & &  \mbox{for }  \yaming { t > v_0/\mu } \mbox{ and } v_t \not\in [v^-(t),v^+(t)],
\end{array}
\right.
\end{equation}
where the limits of the velocity interval are defined by
\begin{equation}\label{ga}
v^-(t)=v_0-\mu t,\qquad v^+(t)=\big(\sqrt{v_0}-\sqrt{\mu t}\big)^2 .
\end{equation}
This shows that, if the endpoint $(v_t,t)$ lies in the area bounded by $v^-(s)$ and $v^+(s)$ with
$s>v_0/\mu$, as shown in Fig.~\ref{FigDryonlyPathPlane}, then the optimal path is an indirect path, otherwise it is a direct path. Intermediate paths do not enter in this result, but will be useful when we treat the Jacobian.

\begin{figure}
\begin{center}
\includegraphics[scale=0.5]{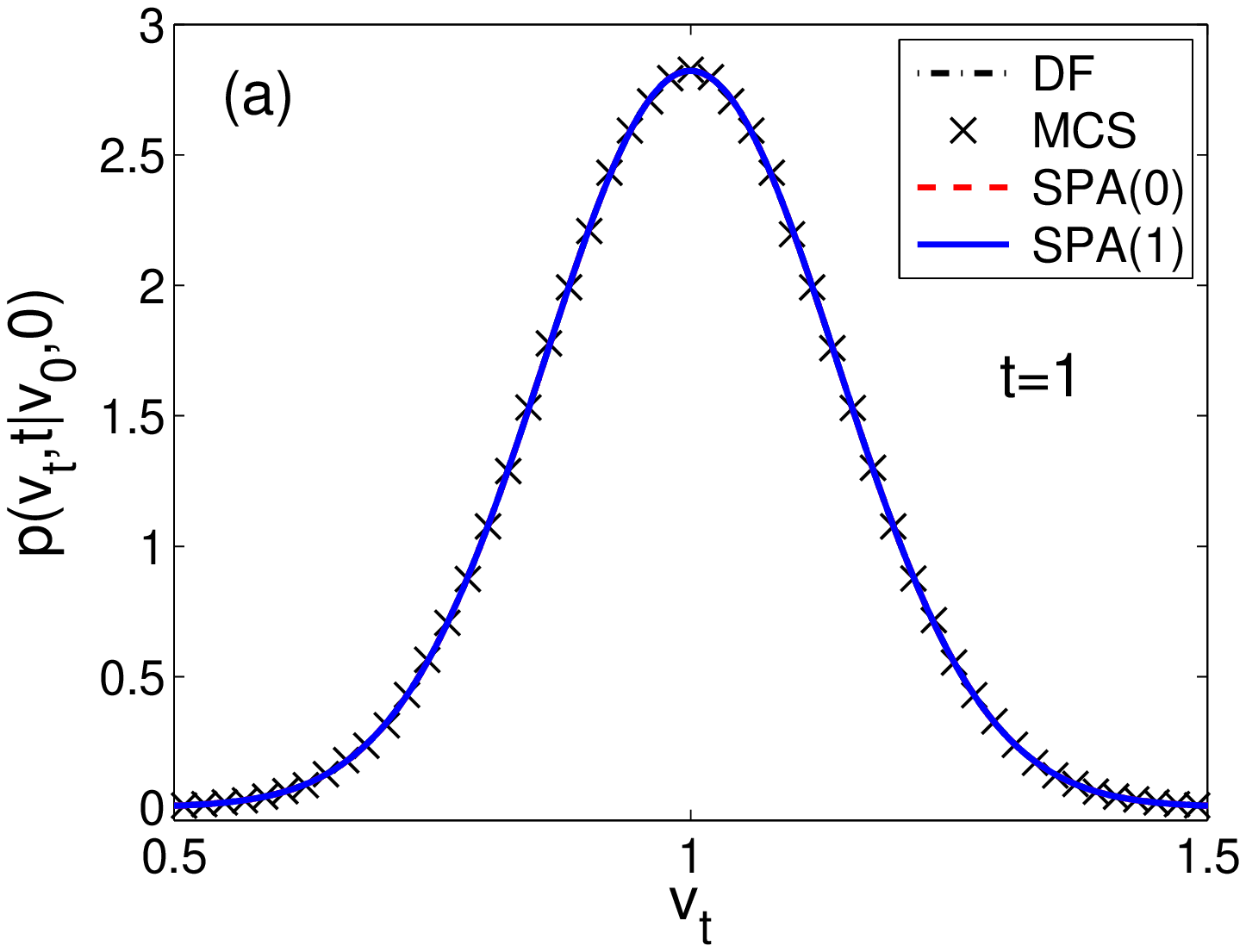}
\includegraphics[scale=0.5]{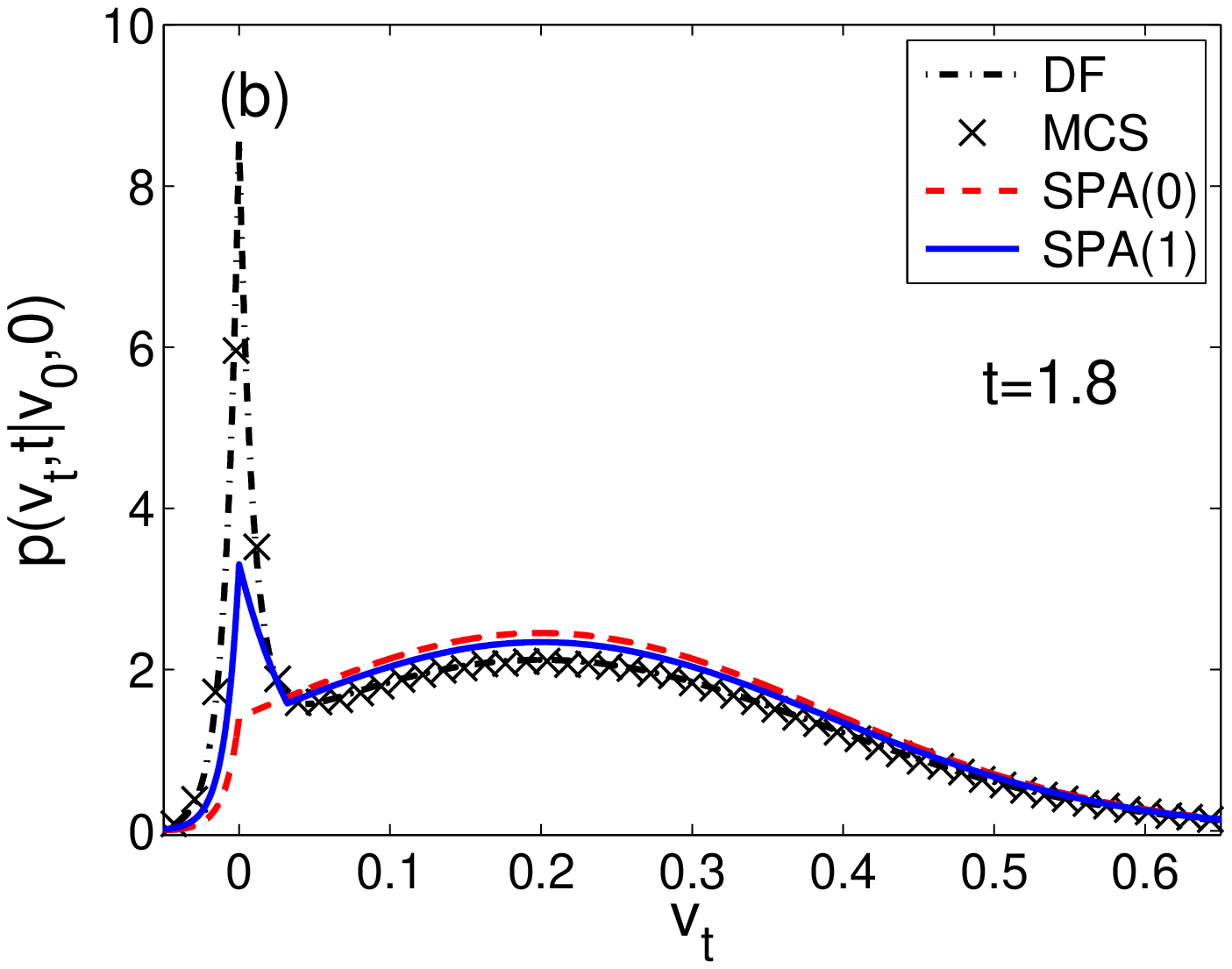}\\
\includegraphics[scale=0.5]{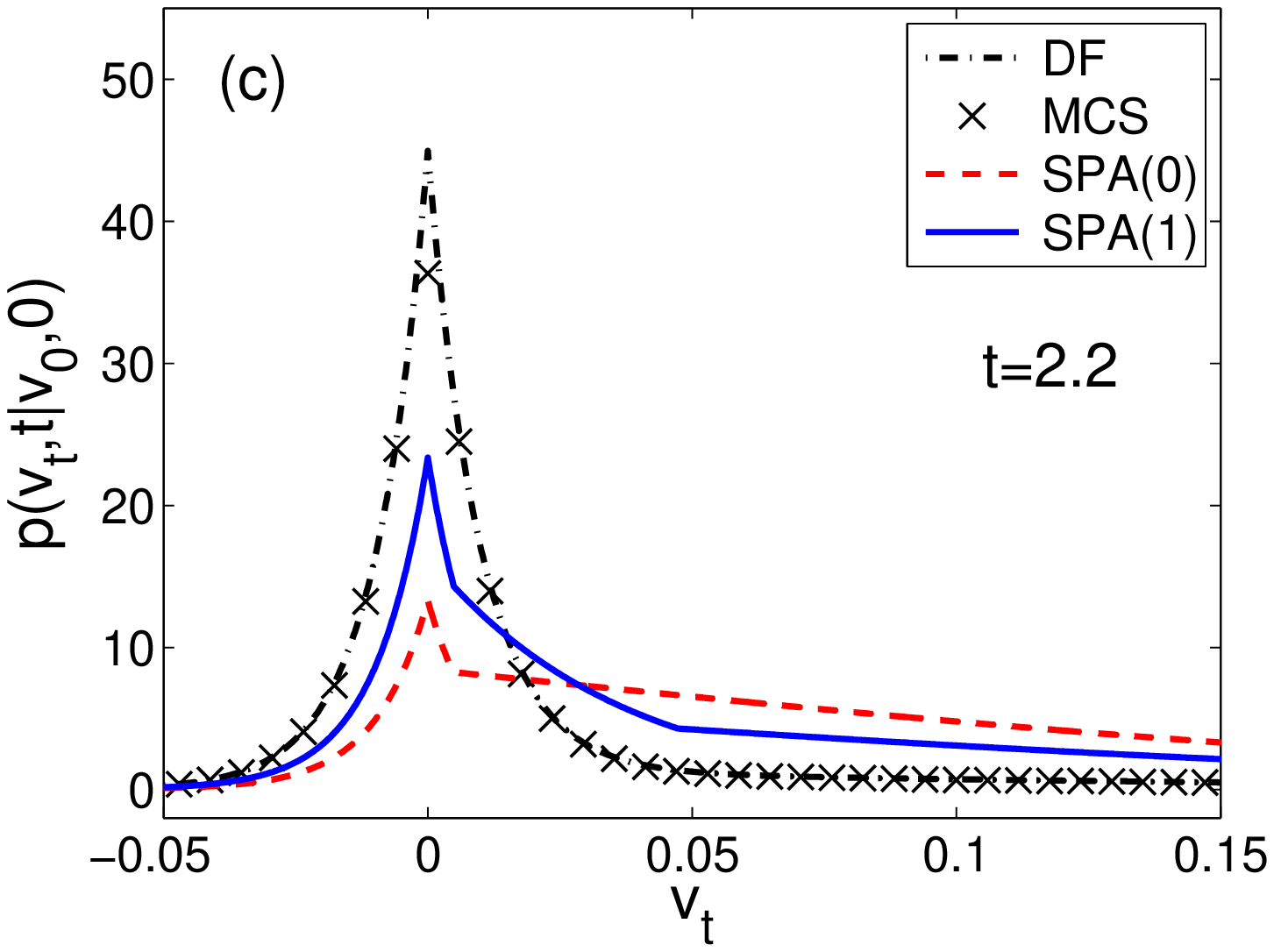}
\includegraphics[scale=0.5]{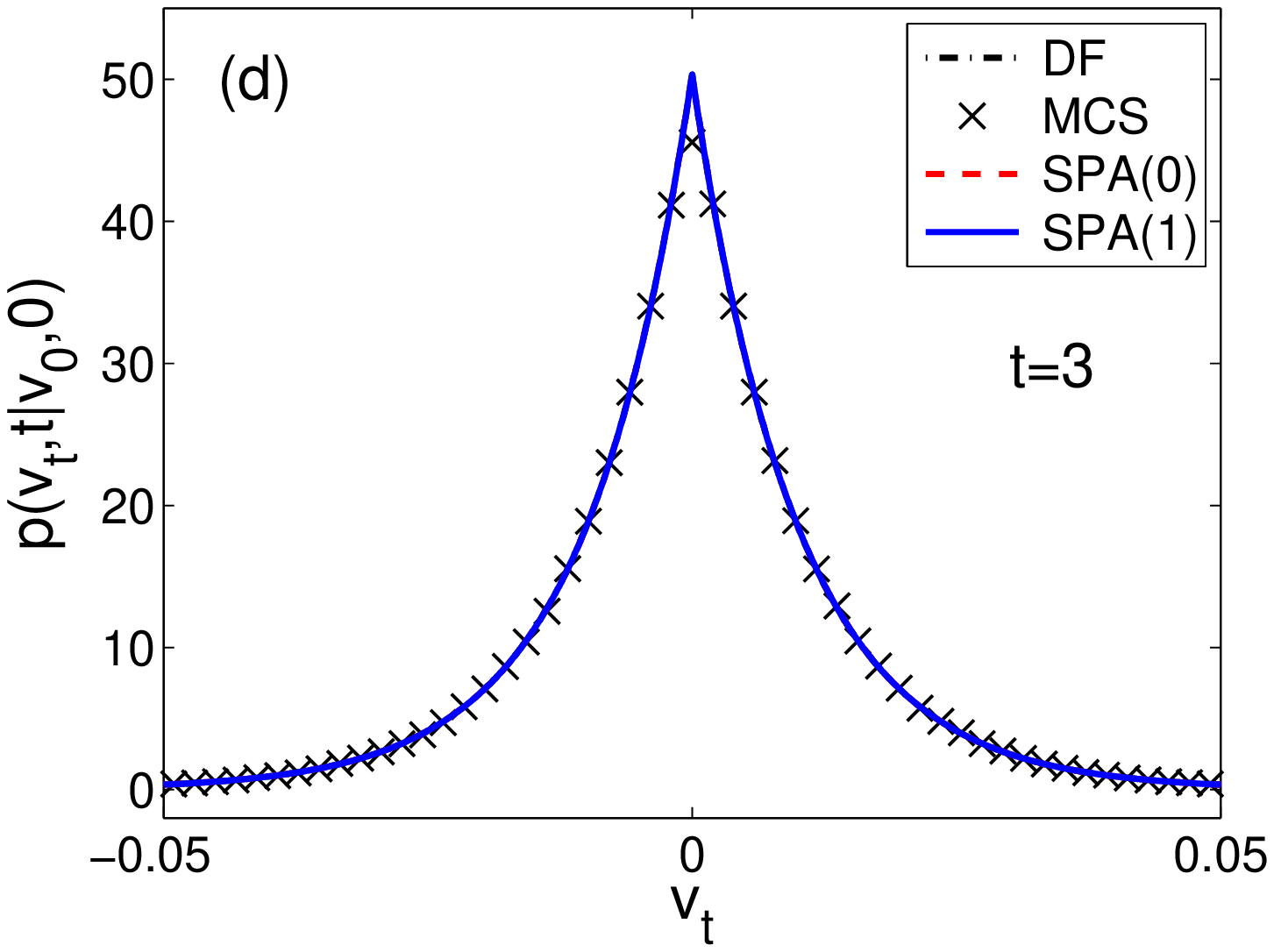}
\end{center}
\caption{(Color online) Propagator of the dry friction model (\ref{EqLangevinOnlyDryFriction}) for initial velocity $v_0=2$, $\mu=1$, $D=0.01$ and different values of time: (a) $t=1$, (b) $t=1.8$, (c) $t=2.2$, and (d) $t=3$. Here DF denotes the exact analytical result (\ref{EqanalyticalSolutionDryOnly}), MCS the Monte Carlo simulation of Eq.~(\ref{EqLangevinOnlyDryFriction}) with step size $ 0.0001 $ and an ensemble of $10^6$ realizations, SPA(0) the leading-order saddle-node expansion [see Eqs.~(\ref{EqPropagatorZero1}) and (\ref{EqPropagatorZero2}) in Sec.~\ref{subsecDryonly0})], and SPA(1) the higher-order saddle-node expansion [see Eqs.~(\ref{dryDirectAction}), (\ref{dryIntermediateAction}),  and (\ref{dryIndirectAction}) in
Sec.~\ref{subsecDryonly1})].}
\label{FigDrySPAanalyticJacobian}
\end{figure}

The SPA(0) of the propagator is obtained from this result by substituting the corresponding action in (\ref{eqspa0}). For $t<v_0/\mu$ we find
\begin{equation}\label{EqPropagatorZero1}
p^{(0)}(v_t,t|v_0,0)=
N_1\left\{
\begin{array}{lll}
\exp[-(v_t-v_0+\mu t)^2/(4Dt) -\mu v_t/D] & & \mbox{for } v_t < 0\\
\exp[-(v_t-v_0+\mu t)^2/(4Dt)] & & \mbox{for } v_t>0,
\end{array}
\right.
\end{equation}
whereas for $t>v_0/\mu$ we find
\begin{equation}
\label{EqPropagatorZero2}
p^{(0)}(v_t,t|v_0,0)=
 N_2\left\{
\begin{array}{lll}
\exp[-(v_t-v_0+\mu t)^2/(4Dt)- v_t/D] & & \mbox{for } v_t < v^-(t)\\
\exp[-(v_t-v_0+\mu t)^2/(4Dt)] & & \mbox{for } v_t > v^+(t)\\
\exp(-\mu |v_t|/D) & &  \mbox{for }  v_t\in [v^-(t),v^+(t)],
\end{array}
\right. 
\end{equation}
where $ N_1 $ and $ N_2 $ are normalization factors. This result is compared in Fig.~\ref{FigDrySPAanalyticJacobian} with the exact propagator. There we see that the SPA(0) is a good approximation of $p(v,t|v_0,0)$ at short and long times, but does not capture the bimodal structure of the propagator arising when the optimal path hits the origin for times close to $t=v_0/\mu$. While a kink of the exact propagator shows up at time $ t<v_0/\mu $ in the region $ v_t>0 $ [see Fig.~\ref{FigDrySPAanalyticJacobian}(b)], the corresponding kink of the SPA(0) only appears at time $ t>v_0/\mu $ [see Fig.~\ref{FigDrySPAanalyticJacobian}(c) as well as Eqs.~(\ref{EqPropagatorZero1}) and (\ref{EqPropagatorZero2})]. For comparison, we also show in Fig.~\ref{FigDrySPAanalyticJacobian} the result of
the propagator obtained from Langevin simulations of the SDE using the standard Euler-Maruyama
integration scheme. The application of this scheme is stable for the piecewise-smooth SDE and only
requires that we choose the integration time step small enough so that we can reproduce the cusp seen
in its propagator. 

We will see in the next section that the inclusion of the Jacobian in the saddle-point approximation enables us to reproduce this bimodality more accurately. An important remark before we get to that part is that the SPA(0) yields the same approximation as that obtained from the Freidlin-Wentzell (FW) large-deviation theory of dynamical systems perturbed by noise \cite{freidlin1984}. Indeed, it is not difficult to verify that the limit
\begin{equation}
I(v,t|v_0,0)=\lim_{D\ra 0} -4D\, \ln p(v,t|v_0,0) = \lim_{D\ra 0} -4D\, \ln 
\hat p\left(\frac{\mu}{D} v_t,\frac{\mu^2}{D}t\,\right|\!\left.
\frac{\mu}{D}v_0,0\right),
\end{equation}
which defines in the FW theory the so-called pseudo- or quasipotential $I(v,t|v_0,0)$, is equal to the SPA(0) action. This shows that the application of the FW theory to nonsmooth SDEs can be done following the heuristic principles formulated above. In this case, the different optimal paths that we have found can be associated, in the low-noise limit, with two different physical modes of motion of the noiseless system: Direct paths represent slip motion, whereas indirect paths represent stick motion \cite{baule2010,baule2011}. For any fixed value of $v_t$, the optimal path will always be an indirect path if the time is sufficiently large, which means that all optimal paths are indirect paths in the limit $t\to\infty$.

\subsection{Corrected action}
\label{subsecDryonly1}

One way to correct the SPA(0) is to use the optimal paths obtained before but to evaluate their action by including the Jacobian in the action:
\begin{equation}\label{EqfirstOrderActionDryonly}
S[v]=\int_0^t\left\{[\dot{v}+\mu \sigma(v)]^2-4D\mu\delta(v)\right\}\, \mathrm{d}s.
\end{equation}
This defines a first-order saddle-point approximation [SPA(1)] of the propagator that retains the optimal paths of SPA(0) but includes the subdominant correction of the Jacobian, which is multiplied by $D$.

To obtain this approximation, we need to evaluate how the contribution arising
from the $\delta$ function in Eq.~(\ref{EqfirstOrderActionDryonly}) changes the action for the different paths considered.
For direct paths, there is obviously only a contribution when the $v=0$ axis is crossed, so that
\begin{equation}
\label{ra}
\int_0^t\delta(v_d^{(0)})\mathrm{d}s=
\left\{
\begin{array}{lll}
t/(v_0-v_t) & & \mbox{for } v_t < 0\\
0 & & \mbox{for } v_t>0 .\\
\end{array}
\right.
\end{equation}
Thus, the corresponding corrected action is 
\begin{equation}
S[v_d^{(0)}] =
\left\{
\begin{array}{lcl}
(v_t-v_0+\mu t)^2/t-4\mu v_t-4D\mu t/(v_0-v_t) &\mbox{ for }& v_t< 0 \\
(v_t-v_0+\mu t)^2/t & \mbox{ for } & v_t> 0. 
\end{array}
\right. \label{dryDirectAction}
\end{equation}

For intermediate paths, the evaluation of the Jacobian term is straightforward as well.
Any sensible representation of the $\delta$ function will result in a symmetric average of the inverse slopes of the path, leading to
\begin{equation}
\label{rb}
\int_0^t\delta(v_{int}^{(0)}) d s =
\frac{1}{2\big|\dot{v}^{(0)}_{int}(t_0^-,t_0)\big|}+\frac{1}{2\big|\dot{v}^{(0)}_{int}(t_0^+,t_0)\big|}
= \frac{t}{v_0+|v_t|},
\end{equation}
where we have used the condition $t_0=v_0t/(v_0+|v_t|)$. 
\yaming{
Hence it follows from Eqs.~(\ref{EqfirstOrderActionDryonly}) and (\ref{rb}) that 
\begin{equation}
\label{dryIntermediateAction}
S[v_{int}^{(0)}] = (|v_t|+v_0-\mu t)^2/t+4\mu |v_t|-4D\mu t/(v_0+|v_t|). 
\end{equation} 
}

Indirect paths require a closer inspection: These paths vanish over an entire interval of time, so the contribution originating from the Jacobian is ill defined. To treat this, we follow the previous principle that paths on the attractor do not contribute to the action and posit the following two additional heuristic principles: (iii) Parts of indirect paths on the attractor are not considered as contributing to the Jacobian and (iv) nonvanishing parts of indirect paths contribute, as for intermediate paths, to the Jacobian in a weighted average way.
With these principles, the corrected action of indirect paths is finite and equal to 
\begin{equation}
\label{rc}
\int_0^t\delta(v_{ind}^{(0)}) \mbox{d}s =
\frac{1}{2\big|\dot{v}_{ind}^{(0)}(t_1^-,t_1,t_2)\big|}+\frac{1}{2\big|\dot{v}_{ind}^{(0)}(t_2^+,t_1,t_2)\big|}
= \mu,
\end{equation}
where we have used the conditions $ t_1=v_0/t $ and $ t_2=t-|v_t|/\mu $.
Thus the corresponding action (\ref{EqfirstOrderActionDryonly}) can be evaluated as 
\begin{equation}
S[v_{ind}^{(0)}] = 4\mu |v_t|-4D\mu \qquad \mbox{ for }       t>(v_0+|v_t|)/\mu. \label{dryIndirectAction}\\
\end{equation}

By properly comparing the corrected actions (\ref{dryDirectAction}), (\ref{dryIntermediateAction}), and (\ref{dryIndirectAction}), we can determine which path is minimal depending on $v_t$ and $t$ to obtain the propagator $p^{(0;1)}(v,t|v_0,0)$ [see Eq.~(\ref{EqfirstOrderPropagatorZeroOrderPath})].
The result of this minimization is shown in Fig.~\ref{FigDrySPAanalyticJacobian} as SPA(1).
We see that the inclusion of the Jacobian correction qualitatively improves the
propagator as compared to the lowest-order approximation,
SPA(0) of Sec.~\ref{subsecDryonly0}, even though there are still some deviations in the transient regime where the exact propagator shows a bimodality. In particular, at time $ t<v_0/\mu $ the SPA(1) already shows the kink corresponding to that of the exact propagator in the region $ v_t>0 $ [see Fig.~\ref{FigDrySPAanalyticJacobian}(b)]. This phenomenon is different with that of the SPA(0), which does not have such a kink at $ t<v_0/\mu $. We also observe in Fig.~\ref{FigDrySPAanalyticJacobian}(c) that the SPA(1) develops two kinks in the region $ v_t>0 $ in an intermediate time interval. These kinks are caused by the intermediate path, which becomes the optimal path for certain choices of the end points. The SPA(0), by contrast, has only one kink in the positive region, since for this approximation intermediate paths can never be the optimal path [see Eq.~(\ref{EqzeroOrderPathDryonly})]. 
These kinks are a well-known artifact of low-noise approximations 
and are smoothed in the exact propagator by the finite diffusion.

\subsection{Corrected action with corrected path}
\label{subsecSPAFirstOrderActionFirstOrderPath}

The SPA(1) corrects the SPA(0) by including the Jacobian term in the action, while using the optimal paths of SPA(0), i.e., the paths that minimize the zeroth-order action $S^{(0)}[v]$. As a further correction to this approximation, it is tempting to obtain the optimal paths by minimizing the corrected action $S[v]$ with the Jacobian, thereby constructing a ``full'' first-order approximation. 

Unfortunately, this approach does not work as the action turns out to diverge for $v_t=0$. To see this, evaluate the
action of Eq.~(\ref{EqfirstOrderActionDryonly}) for an intermediate path
$ v_{int}^{(0)} $ with a kink at $t_0=t/2$ (see Fig.~\ref{FigpathDryonly}):
\begin{equation}
\label{ac}
S[v_{int}^{(0)}] = 2\mu(|v_t|-v_0)+\mu^2 t+2(v_0^2+v_t^2)/t
-D\mu\left(t/v_0+t/|v_t|\right).
\end{equation}
This value is an upper bound for the minimum of the action,
which determines the density according to Eq.~(\ref{EqfirstOrderPropagatorFirstOrderPath}).
The problem with this result is that $S[v_{int}^{(0)}]\rightarrow -\infty$ when $v_t\rightarrow 0$, leading to
a (nonintegrable) singularity for the propagator $p(v_t,t|v_0,0)$ at $v_t=0$.
Thus the approximation scheme based on obtaining the optimal paths from $S[v]$ results in a 
non-normalizable expression, which implies that a full first-order SPA is not possible 
for the SDE of Eq.~(\ref{EqLangevinOnlyDryFriction}). It is clear that this problem will also arise in any SDE having, as in Eq.~(\ref{EqLangevinOnlyDryFriction}), points where the force of the SDE is discontinuous. One way to approach this problem is to explore regularizations of such discontinuities.

\section{Regularized SDE}
\label{sect.regularization}

As seen in the previous section, the weak-noise
expansion of the path integral
for nonsmooth SDEs faces some
difficulties related to the minimization of the action and the singularity of the Jacobian term.
To treat this problem,
we now consider the regularized SDE of Eq.~(\ref{EqLangevinTanh})
in which
the discontinuous drift $\sigma(v)$ is replaced by the smooth drift
$\tanh\left(v/\varepsilon\right)$ involving the additional (small)
parameter $\varepsilon$. For this smooth SDE, the aforementioned difficulties do not occur: We can minimize the action over smooth
differentiable paths and the Jacobian is well defined. In this section, we are interested in understanding the weak-noise
properties of this regularized SDE.

To investigate the regularized model, we introduce non-dimensional units:
\begin{equation}\label{ca}
u=v/\varepsilon,\qquad \tau=\mu t/\varepsilon .
\end{equation}
Equation (\ref{EqLangevinTanh}) is then simply written as
\begin{equation}\label{EqregularizedLangevin}
\dot{u}=-\tanh (u) +\sqrt{\widetilde{D}}\, \xi(\tau),
\end{equation}
where $\widetilde{D}=D/(\varepsilon\mu)$ is now the only parameter of the model, called the effective diffusion constant.
An important point to note is that the two limits $\varepsilon\rightarrow 0$ and $D\rightarrow 0$ do 
not commute.
In the following, we will be interested in the smooth model for small and
moderate effective noise amplitudes.

As mentioned in the Introduction, 
the propagator of the regularized model does not have a known closed analytic form.
Thus, for benchmarking the weak
noise expansion results, we need to resort to
numerical simulations obtained with the Euler-Maruyama scheme \cite{Kloeden1992NumericalSDE}, which accurately
reproduce, as for the piecewise smooth SDE, all the features of the propagator, provided we choose the integration time step small enough.

As before, the leading order of the weak-noise expansion of Eq.~(\ref{EqregularizedLangevin}) is determined by the action
[see Eqs.~(\ref{EqzeroOrderAction} and (\ref{EqzeroOrderPropagator})]
\begin{equation}\label{EqzeroOrderActionTanh}
S^{(0)}[u]=\int_{0}^{\tau}\left[ \dot{u}(s)+\tanh u(s)\right]^2 \mbox{d}s
\end{equation}
evaluated for the solution of the Euler-Lagrange boundary value problem
[see Eq.~(\ref{EqzeroOrderEL})]
\begin{equation}\label{EqzeroOrderELTanh}
\ddot{u}^{(0)}(s)=\tanh[ u^{(0)}(s)]/\cosh^2[u^{(0)}(s)], \quad
u^{(0)}(0)=u_0, \quad u^{(0)}(\tau)=u_\tau .
\end{equation}
The leading order may be improved by consistently taking
first-order contributions into account. The propagator (\ref{EqfirstOrderPropagatorFirstOrderPath})
is then determined by the action [see Eq.~(\ref{Eqaction})]
\begin{equation}\label{EqfirstOrderActionTanh}
S[u]=\int_{0}^{\tau}\left\{
\left[ \dot{u}(s)+\tanh u(s)\right]^2-2\widetilde{D}/\cosh^2[u(s)] \right\} \mbox{d}s
\end{equation}
evaluated at the first-order path that obeys the boundary value problem (\ref{EqfirstOrderEL})
\begin{equation}\label{EqfirstOrderELTanh}
\ddot{u}^{(1)}(s)=\big(1+2\widetilde{D}\big)\tanh [u^{(1)}(s)]/\cosh^2[u^{(1)}(s)], \quad
u^{(1)}(0)=u_0, \quad u^{(1)}(\tau)=u_\tau .
\end{equation}
The two boundary value problems (\ref{EqzeroOrderELTanh}) and
(\ref{EqfirstOrderELTanh}) differ just by a rescaling of time and can 
be solved using a numerical shooting method (see Appendix \ref{App.pathIntegral}).

Figure~\ref{FigEulerPathEsilon005} compares the numerical Monte Carlo simulations
of Eq.~(\ref{EqregularizedLangevin})
with the zeroth- and first-order weak-noise approximations for small and moderate noise amplitudes.
We see that, while the short- and long-time behaviors are captured quite well by the lowest order
approximation, substantial transient deviations are visible when the maximum of the propagator
approaches the origin. The first-order approximation is in fact able to deal with such a
feature, even for substantial noise amplitudes. Thus the scheme outlined
above can be considered as a candidate to deal with the weak-noise limit even in systems
that mimic the discontinuous drift.

\begin{figure}
\begin{center}
\includegraphics[scale=0.37]{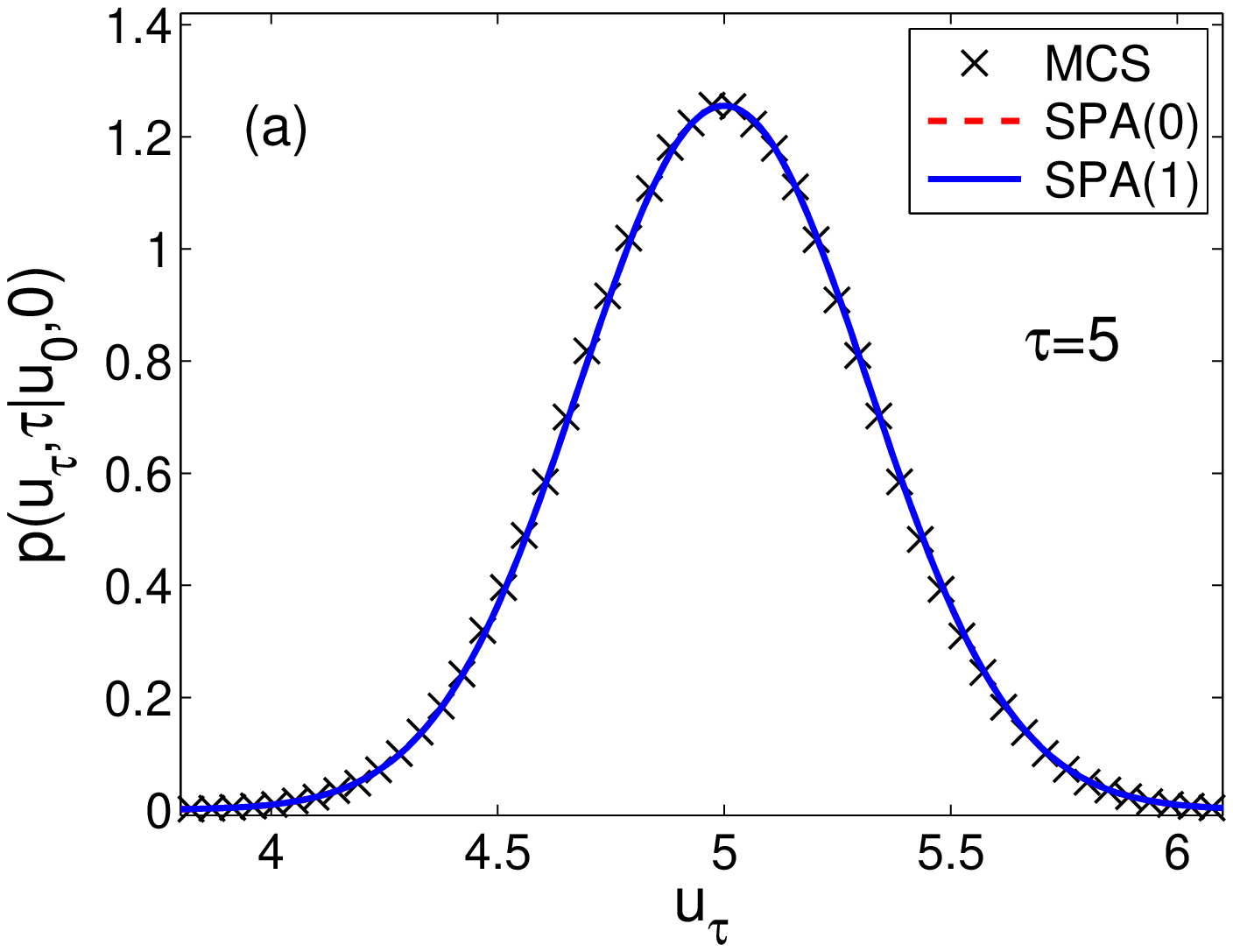}
\includegraphics[scale=0.37]{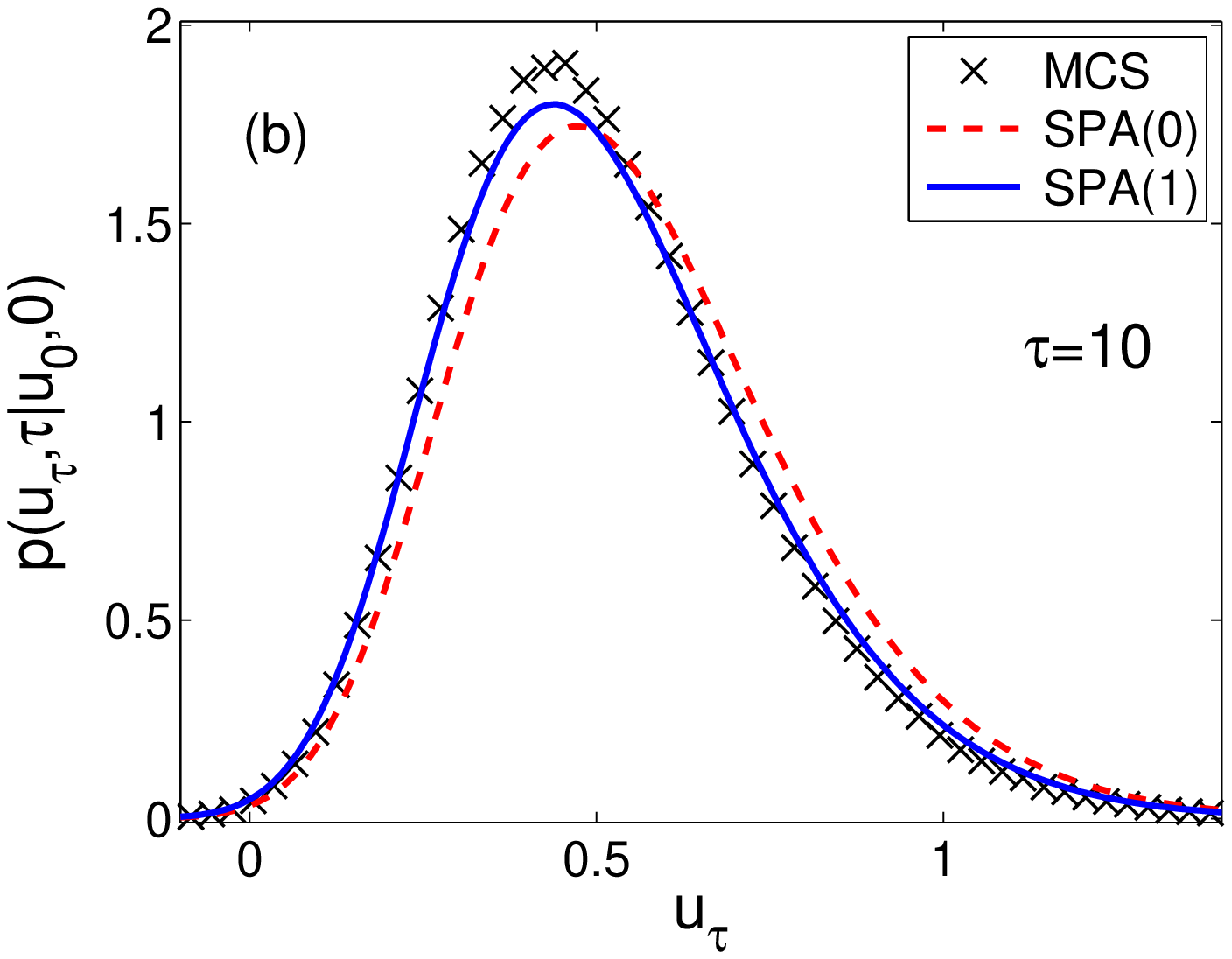}
\includegraphics[scale=0.37]{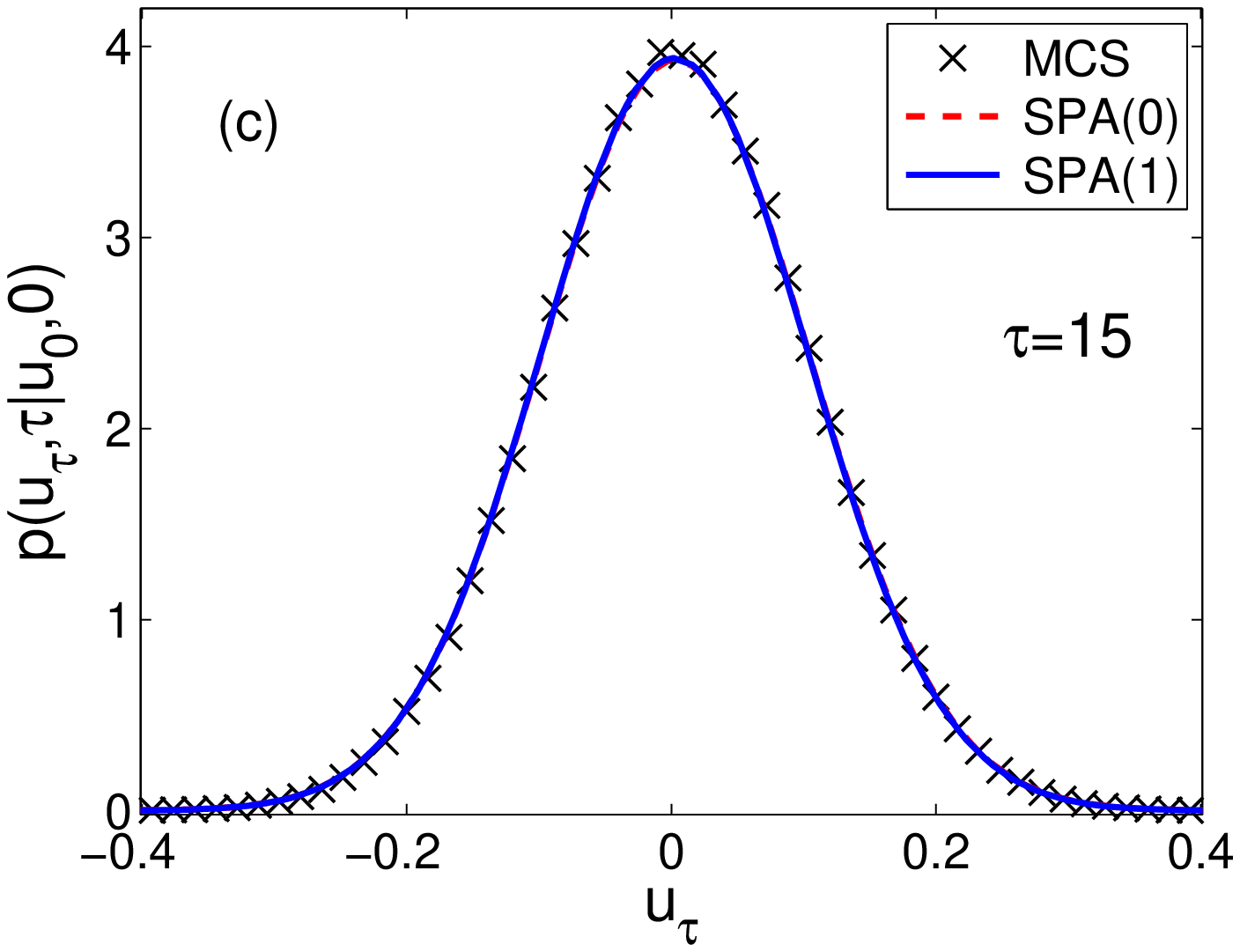}\\
\includegraphics[scale=0.37]{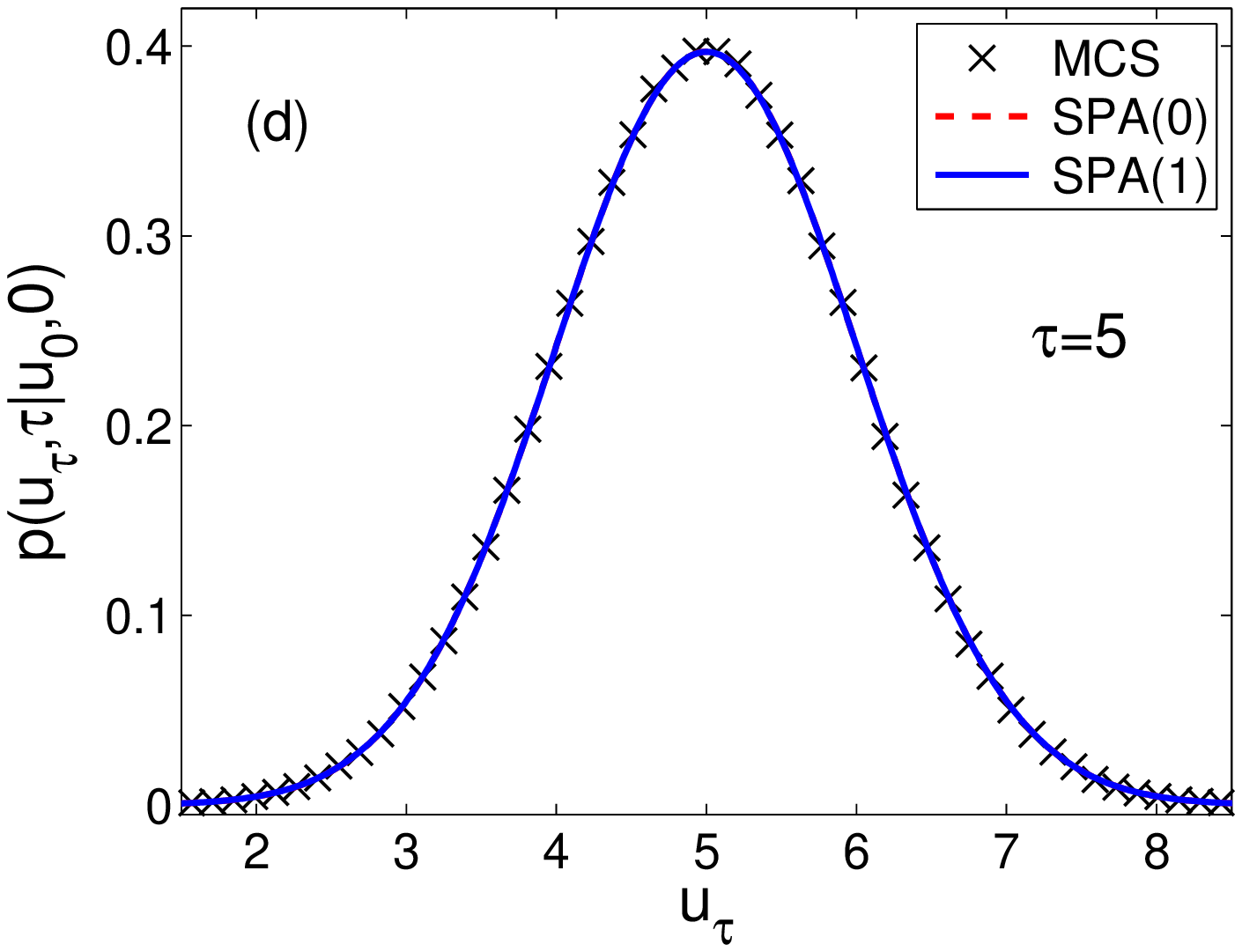}
\includegraphics[scale=0.37]{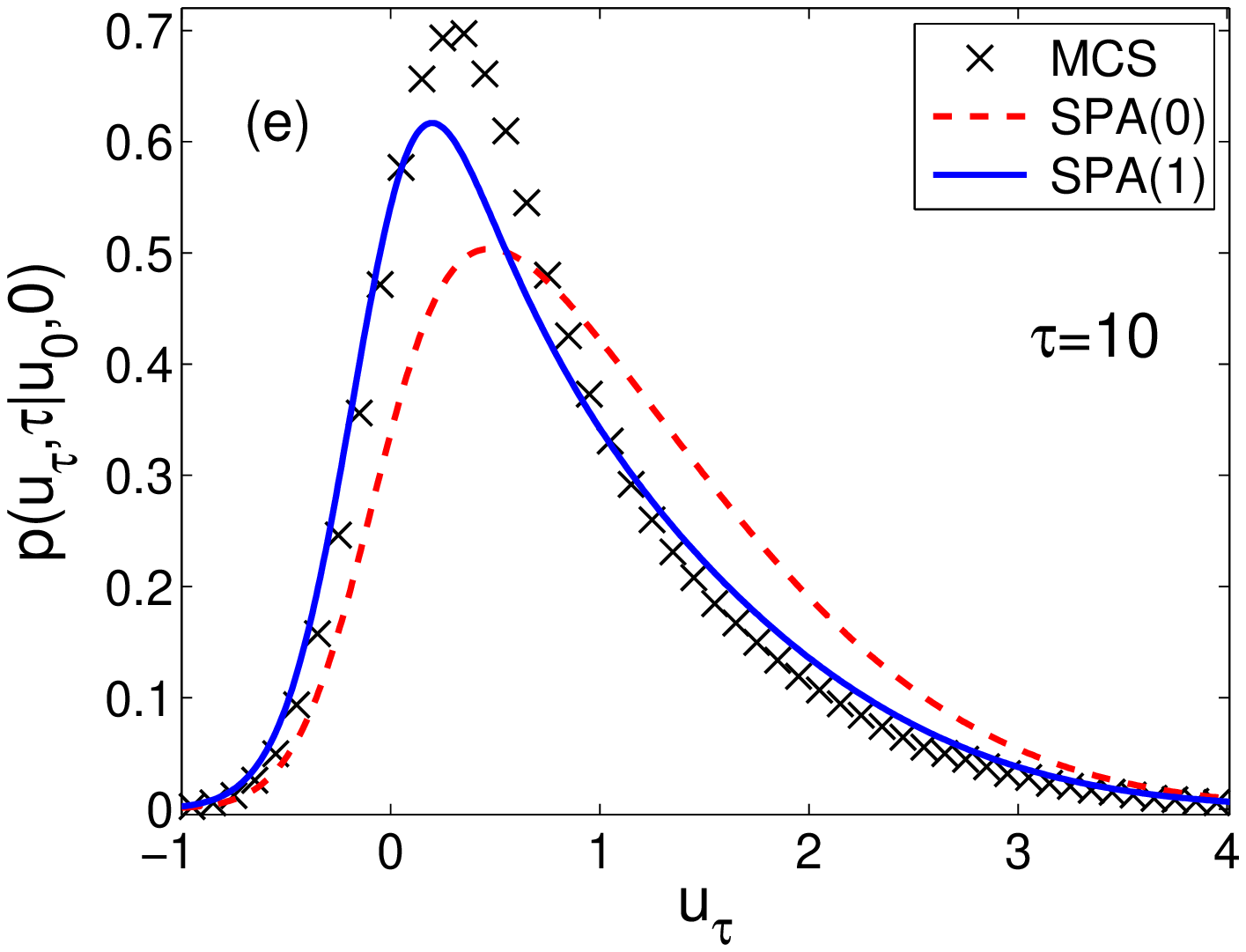}
\includegraphics[scale=0.37]{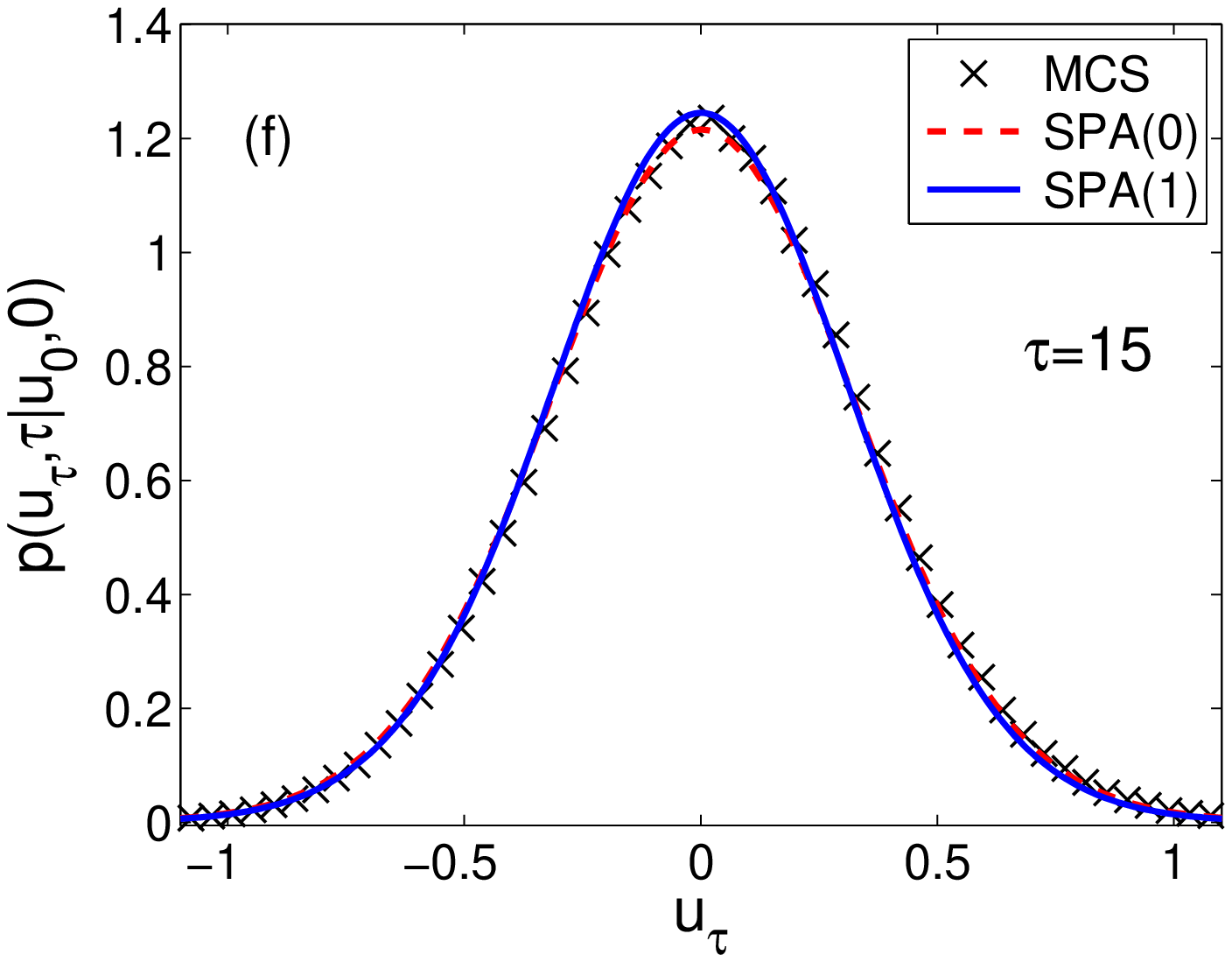}\\
\end{center}
\caption{(Color online) Propagator of Eq.~(\ref{EqregularizedLangevin}) for small and moderate values
of the effective diffusion, (a)--(c) $\widetilde{D}=0.01$ and (d)--(f) $\widetilde{D}=0.1$, initial condition $u_0=10$ and three different values of $ \tau $. Here MCS denotes the Monte Carlo simulations of the Langevin equation (\ref{EqregularizedLangevin}) with time step $ 0.005 $ and an ensemble of $10^6$ realizations, SPA(0) the
lowest order of the saddle node expansion using the action of Eq.~(\ref{EqzeroOrderActionTanh}) [see Eq.~(\ref{EqzeroOrderPropagator})], and SPA(1) the first order of the saddle-node expansion using the action of Eq.~(\ref{EqfirstOrderActionTanh}) [see also Eq.~(\ref{EqfirstOrderPropagatorFirstOrderPath})].}
\label{FigEulerPathEsilon005}
\end{figure}

\begin{figure}
\begin{center}
\includegraphics[scale=0.37]{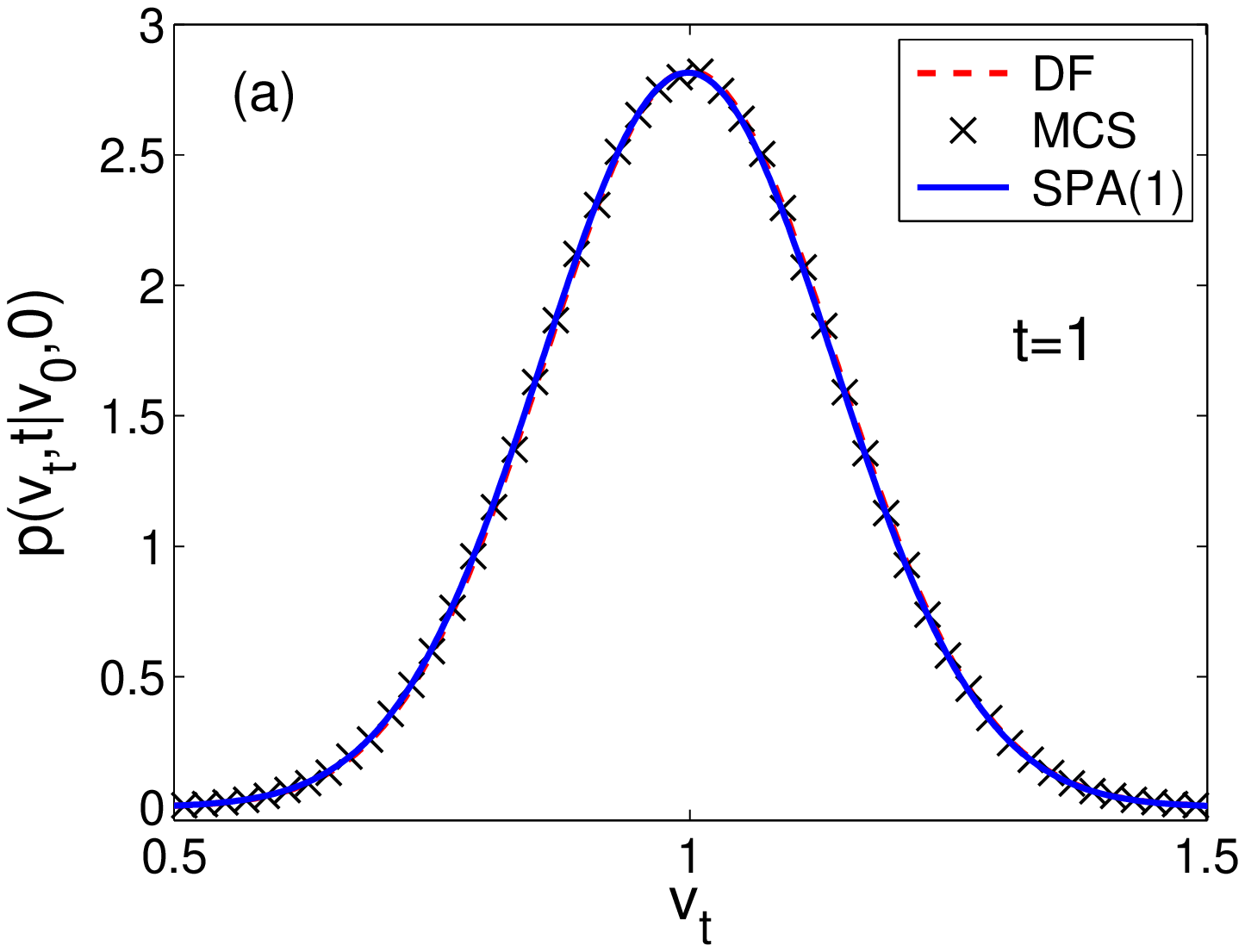}
\includegraphics[scale=0.37]{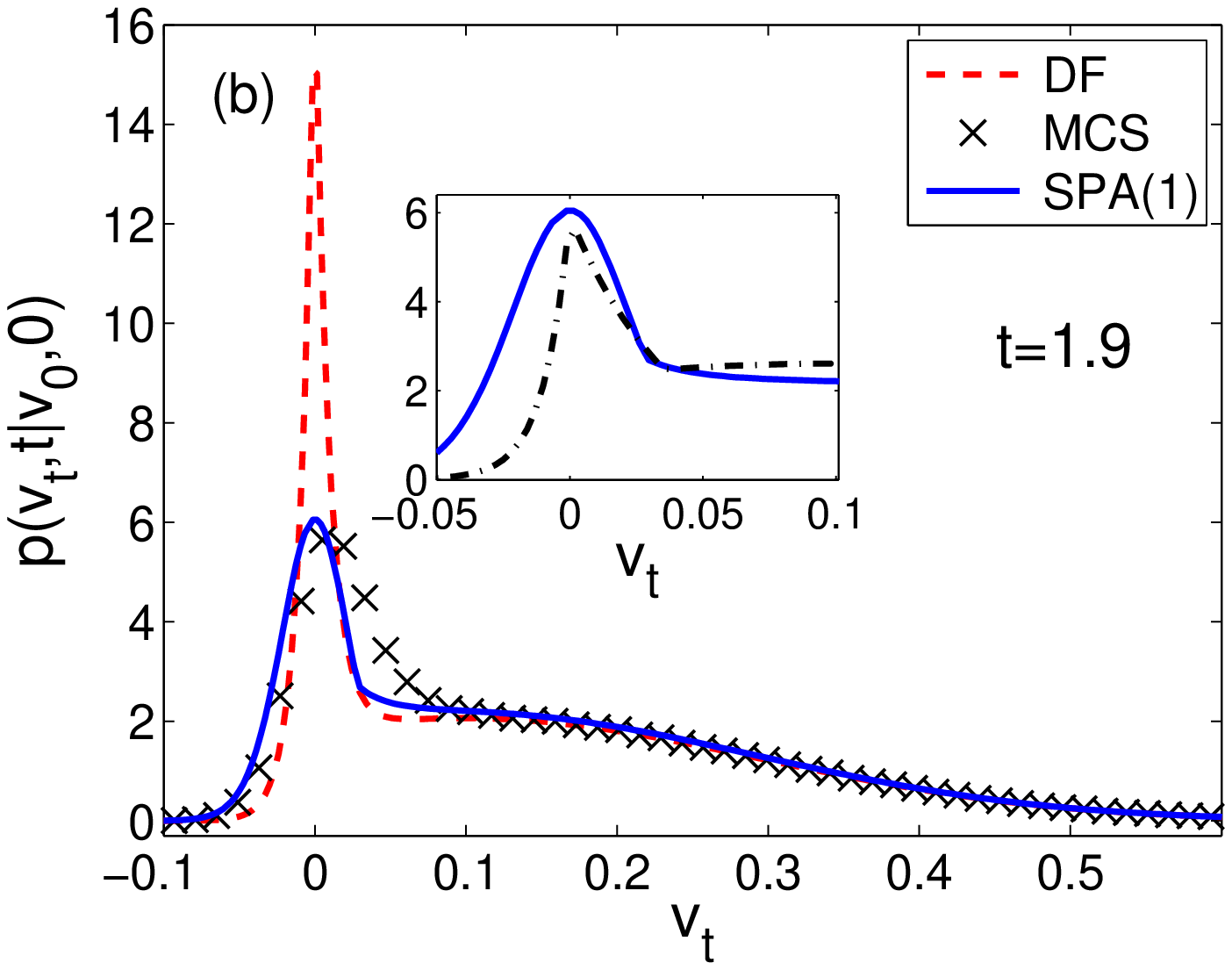}
\includegraphics[scale=0.37]{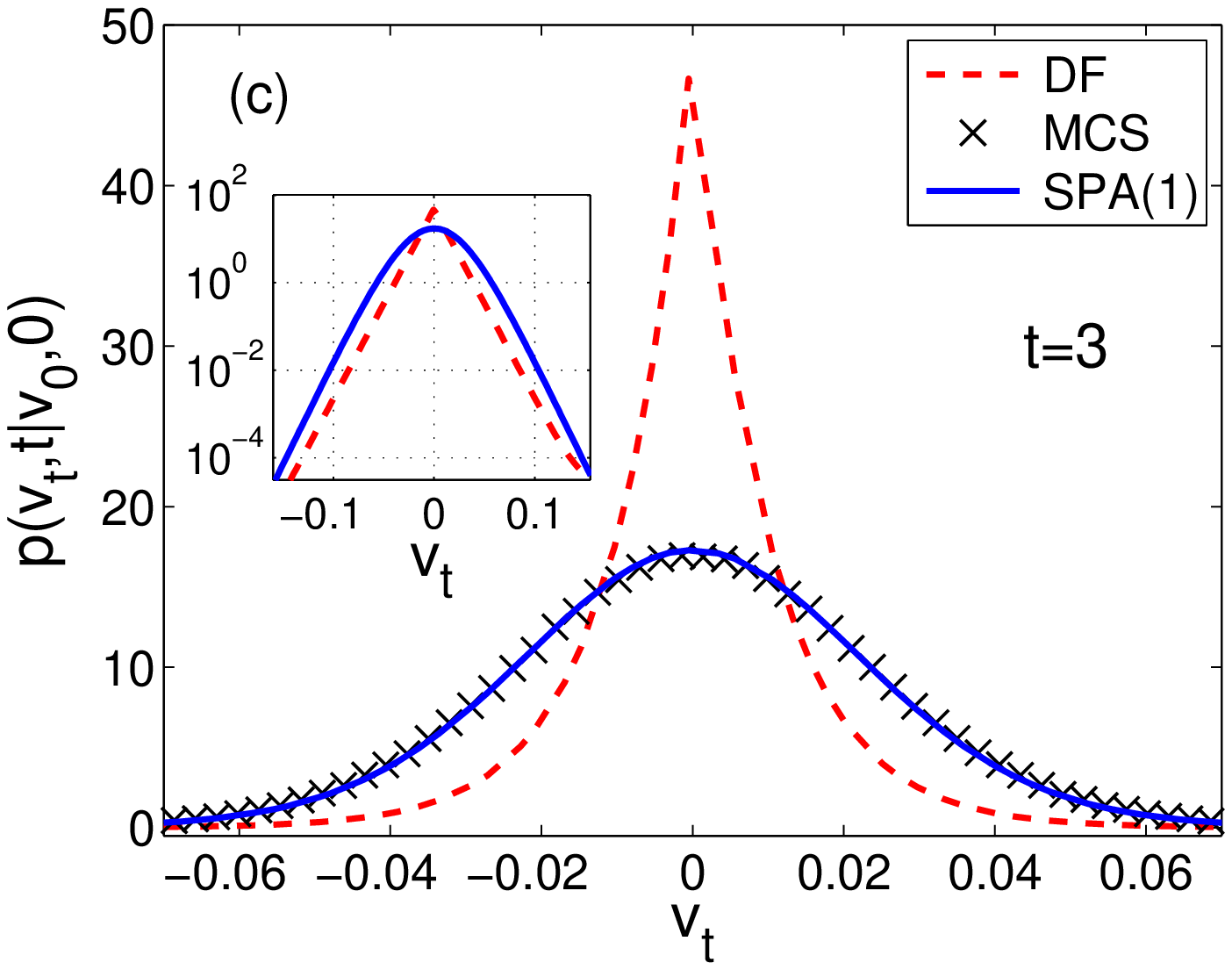}\\
\end{center}
\caption{(Color online) Comparison of the propagators of the dry friction model (\ref{EqLangevinOnlyDryFriction}) and of the regularized scheme (\ref{EqLangevinTanh}) for $\mu=1$, $D=0.01$, $\varepsilon=0.05$, initial condition $v_0=2$, and three different values of time: (a) $t=1$, (b) $t=1.9$, and (c) $t=3$.
Here DF denotes Analytical result for the propagator of the dry friction model [see Eq.  (\ref{EqanalyticalSolutionDryOnly})], 
MCS the Monte Carlo simulation of the Langevin equation (\ref{EqLangevinTanh}) with time step $ 0.005 $ and  an ensemble of $10^6$ realizations, SPA(1) the first order saddle-point approximation scheme using the action of Eq.~(\ref{EqfirstOrderActionTanh}). The inset in (b) shows the SPA(1) of the smooth case (solid line) and the SPA(1) of the dry friction case (dashed line) [see Eqs.~(\ref{dryDirectAction}), (\ref{dryIntermediateAction}), and (\ref{dryIndirectAction})]. Both of these graphs have a kink that matches the structure appearing in the exact solution of the dry friction model (DF). The inset in (c) shows the data for the DF and SPA(1) on a larger scale on a semilogarithmic plot. Large quantitative deviations appear only in a neighborhood of size $ \varepsilon $ of the discontinuity. }\label{FigEulerPathEsilon002}
\end{figure}

Naturally, from our study of the discontinuous model, we cannot expect
the weak-noise approximation to yield the full propagator of the smooth model in
the asymptotic limit $\varepsilon \rightarrow 0$.
However, it should be possible to obtain, at a quantitative level, the main features of the propagator
for suitable small values of $D$ and $\varepsilon$. Of course, an improved
scheme such as the first-order expansion needs to be applied, as one cannot rely on extremely small values
of the effective diffusion to cover cases that are sufficiently close to a discontinuous
drift. Indeed, Fig.~\ref{FigEulerPathEsilon002} shows that this is the case if we translate the results obtained via
Eqs.~(\ref{EqfirstOrderActionTanh}) and (\ref{EqfirstOrderELTanh}) to the dimensional units via
Eq.~(\ref{ca}).
The SPA(1) performs well in short time and is also able to capture the main profiles of the 
piecewise-smooth SDE in moderate and long times. Larger deviations between the discontinuous and the regularized result
appear only in a neighborhood of size $\varepsilon$ of the discontinuity.
Hence we may conclude that, in these particular cases, a suitable regularization and
first-order saddle-point approximation are able to capture the essential features of the piecewise smooth SDE. 

\begin{figure}
\begin{center}
\includegraphics[scale=0.37]{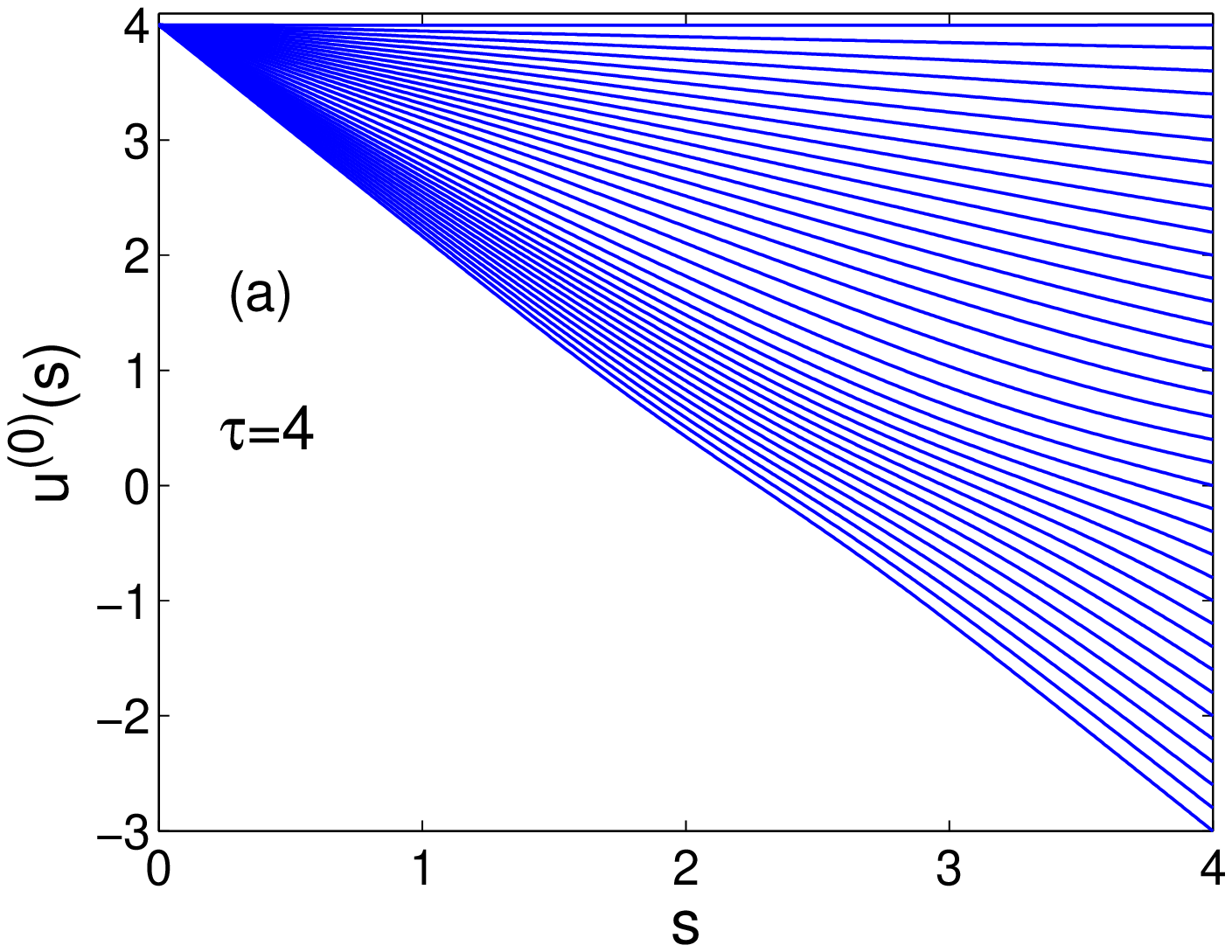}
\includegraphics[scale=0.37]{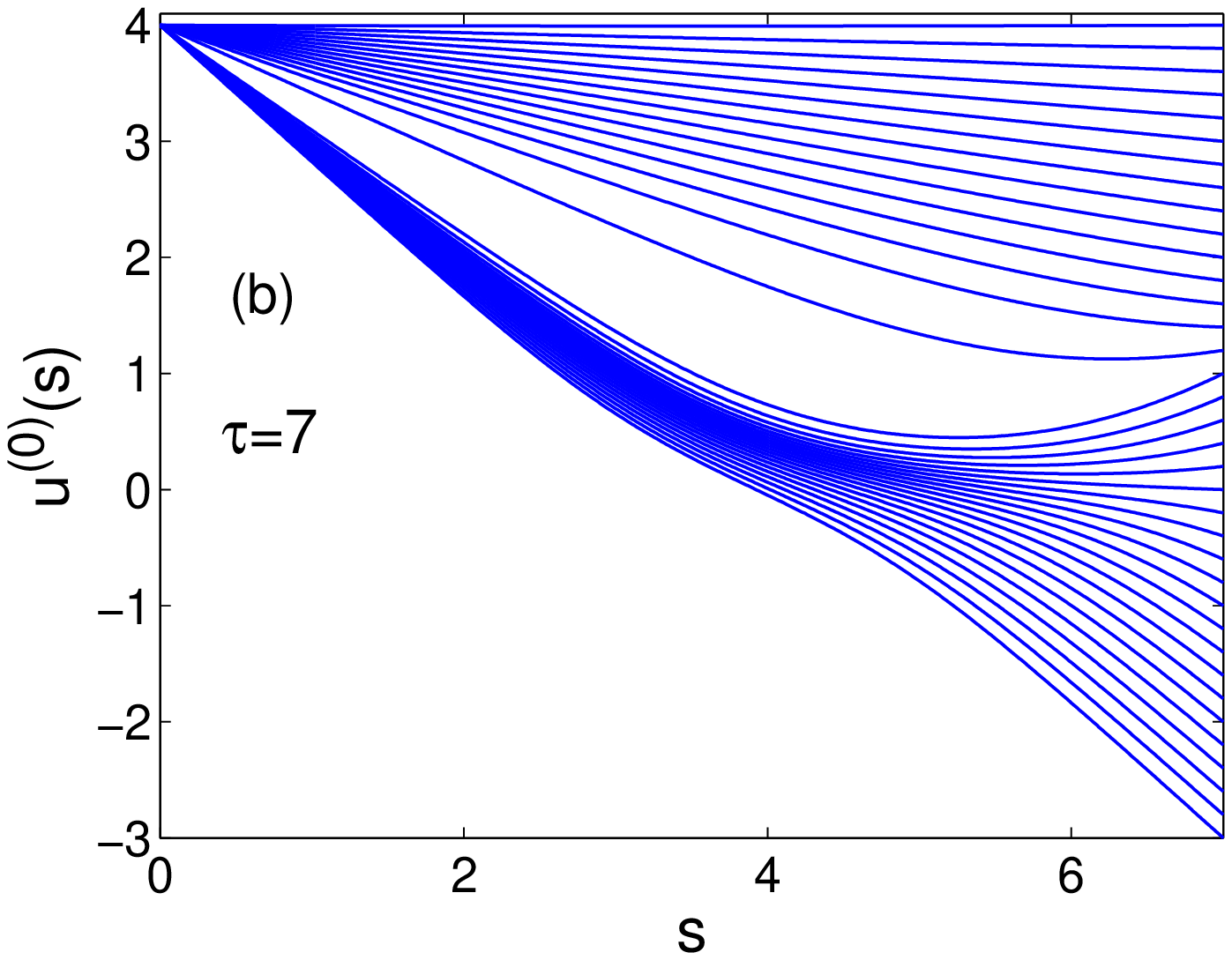}
\includegraphics[scale=0.37]{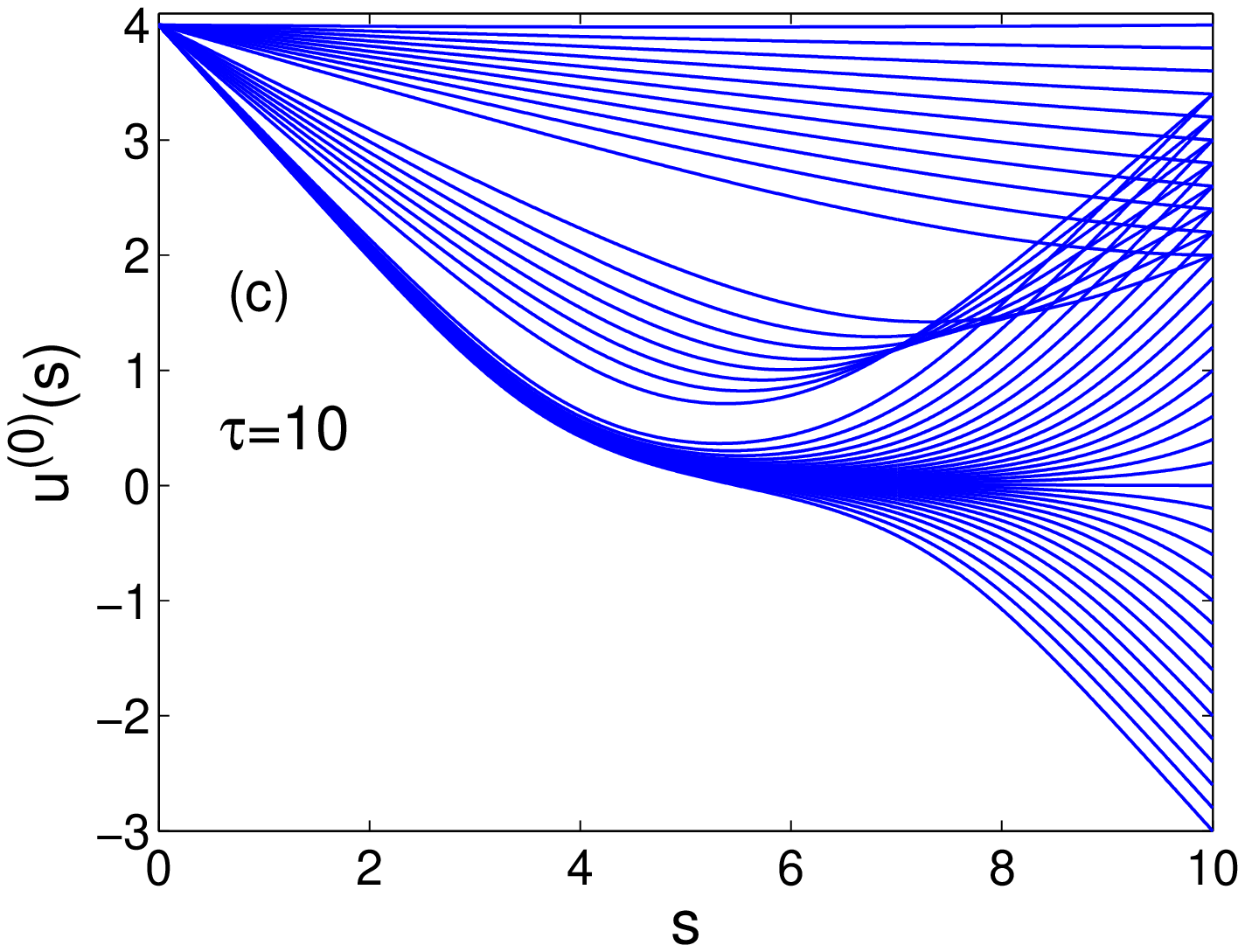}\\
\includegraphics[scale=0.37]{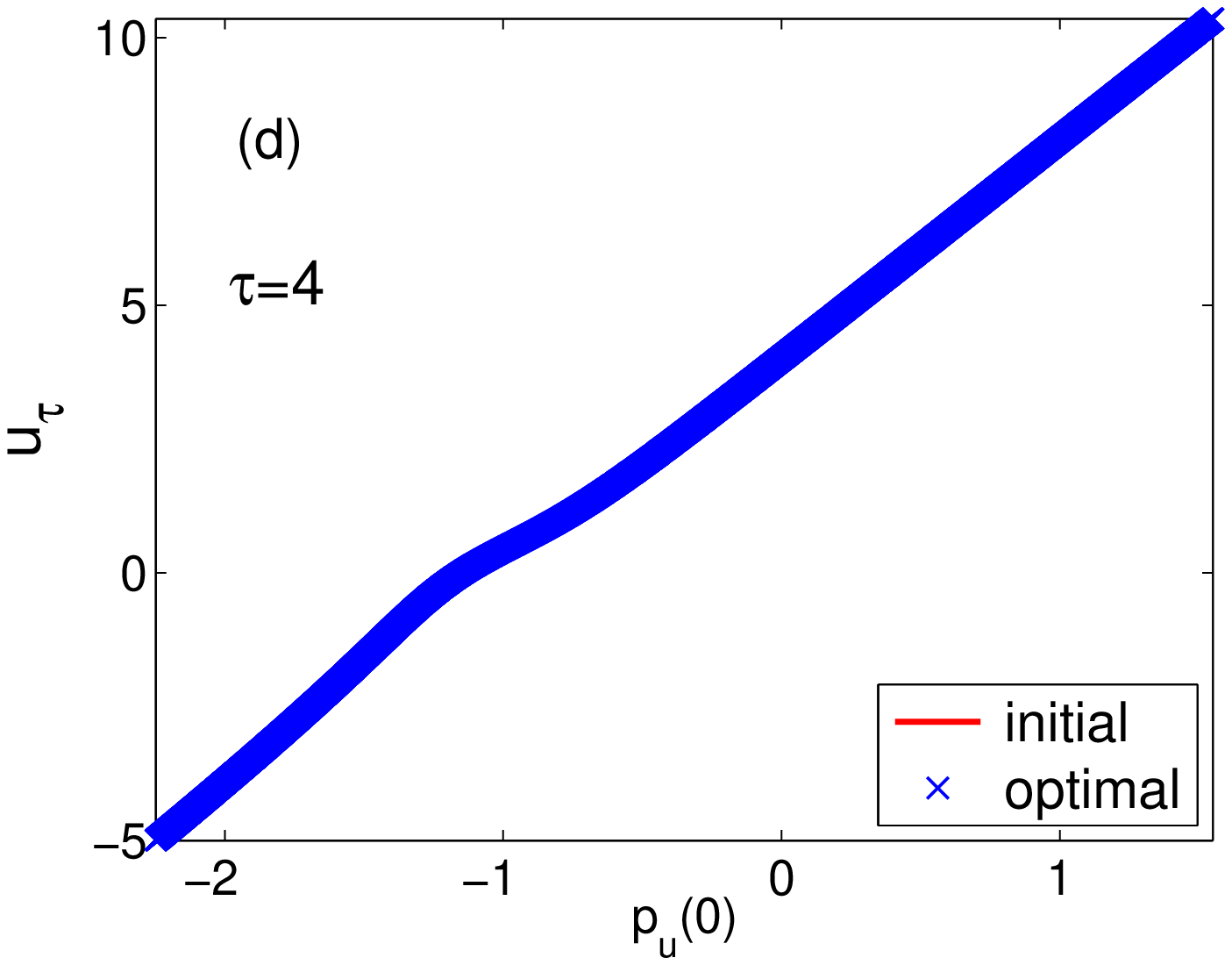}
\includegraphics[scale=0.37]{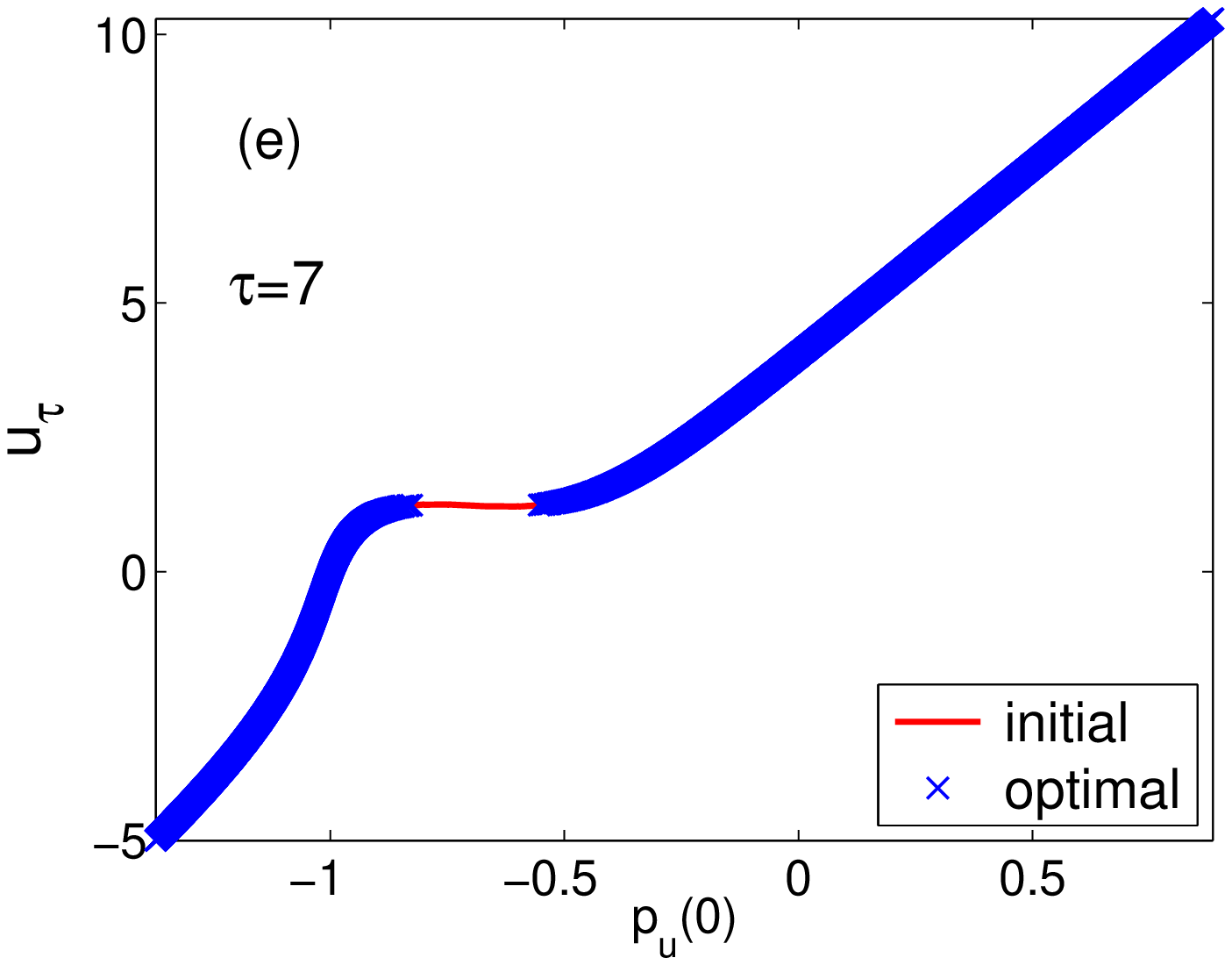}
\includegraphics[scale=0.37]{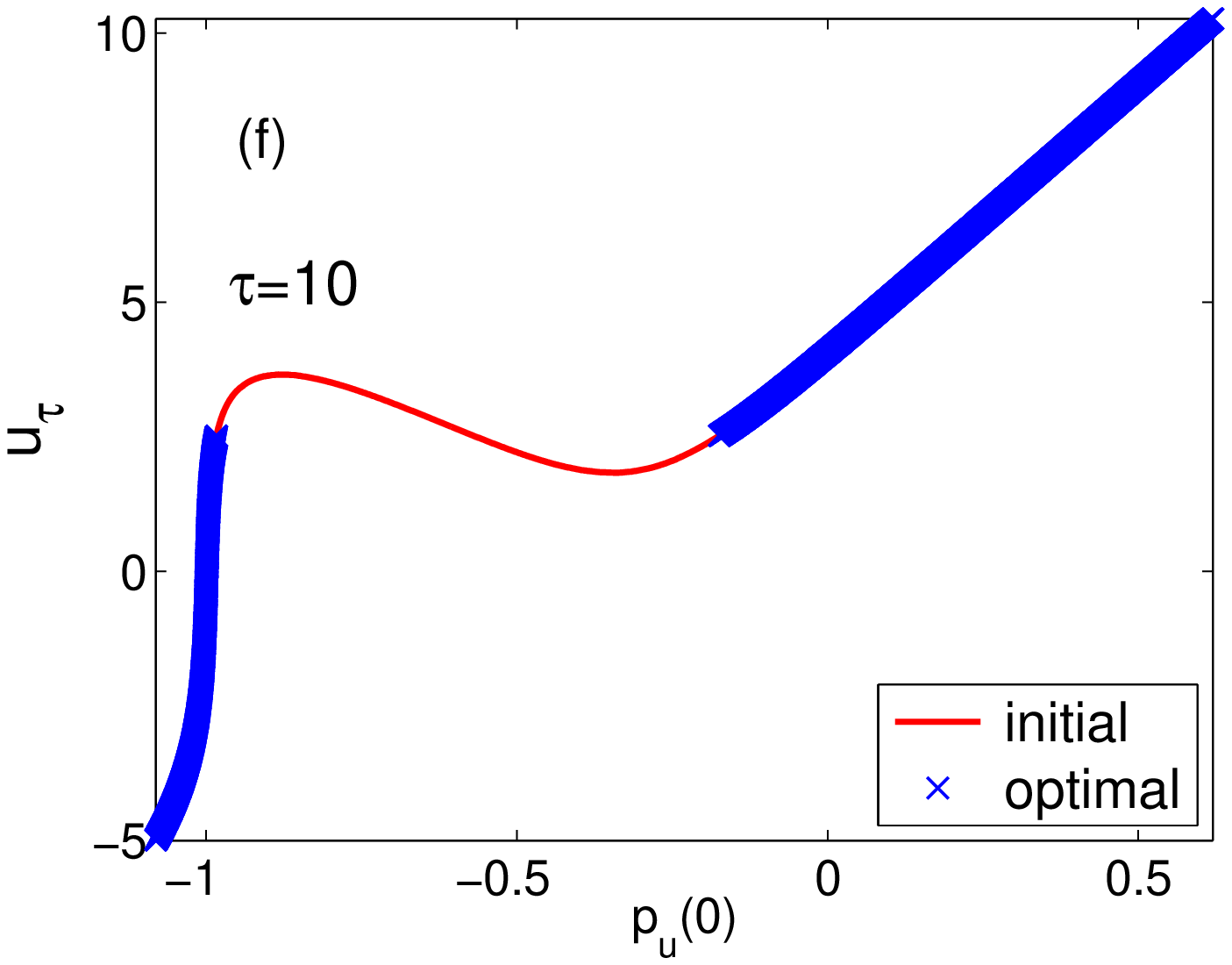}
\end{center}
\caption{(Color online) Solution of the boundary value problem (\ref{EqzeroOrderELTanh}) with the shooting method
for the given value $u_0=4$ and three different values of $\tau$. (a)--(c) Solutions of the
differential equation (\ref{EqzeroOrderELTanh}) for different values of the initial slope
$\dot{u}^{(0)}(0)=p_u(0)$. (d)--(f) Dependence of the final value $u_\tau=u^{(0)}(\tau)$ on the
initial condition $\dot{u}^{(0)}(0)=p_u(0)$ of the differential equation (\ref{EqzeroOrderELTanh}), indicated by the solid line. The symbols indicate the solution that minimizes the
action (\ref{EqzeroOrderActionTanh}).}
\label{FigEndpoints}
\end{figure}

To close this section, let us comment on the kinks appearing in the quasipotential, corresponding to the minimized action. 
These nonanalytic points are well known in large deviation theory to arise from 
the nonuniqueness of the solutions of the Euler-Lagrange boundary value problem \cite{graham1985}.
Since Eqs.~(\ref{EqzeroOrderELTanh}) and (\ref{EqfirstOrderELTanh}) differ only by a rescaling of time,
and since we are here
mainly concerned with the essential structure of the quasipotential we just focus on the
simple zeroth-order expansion (\ref{EqzeroOrderActionTanh}) and (\ref{EqzeroOrderELTanh}).
Figure~\ref{FigEndpoints} shows the solution of the boundary
value problem using the shooting method for a given value of $u_0$. At some finite value of $\tau$,
the boundary value problem develops a cusp singularity, beyond which three
solutions occur in a finite interval of $u_\tau$ values,
corresponding to smooth versions of the previously identified direct, indirect, and intermediate paths.
Hence we recover in the smooth model the same path structure of the discontinuous model.

\begin{figure}
\begin{center}
\includegraphics[scale=0.37]{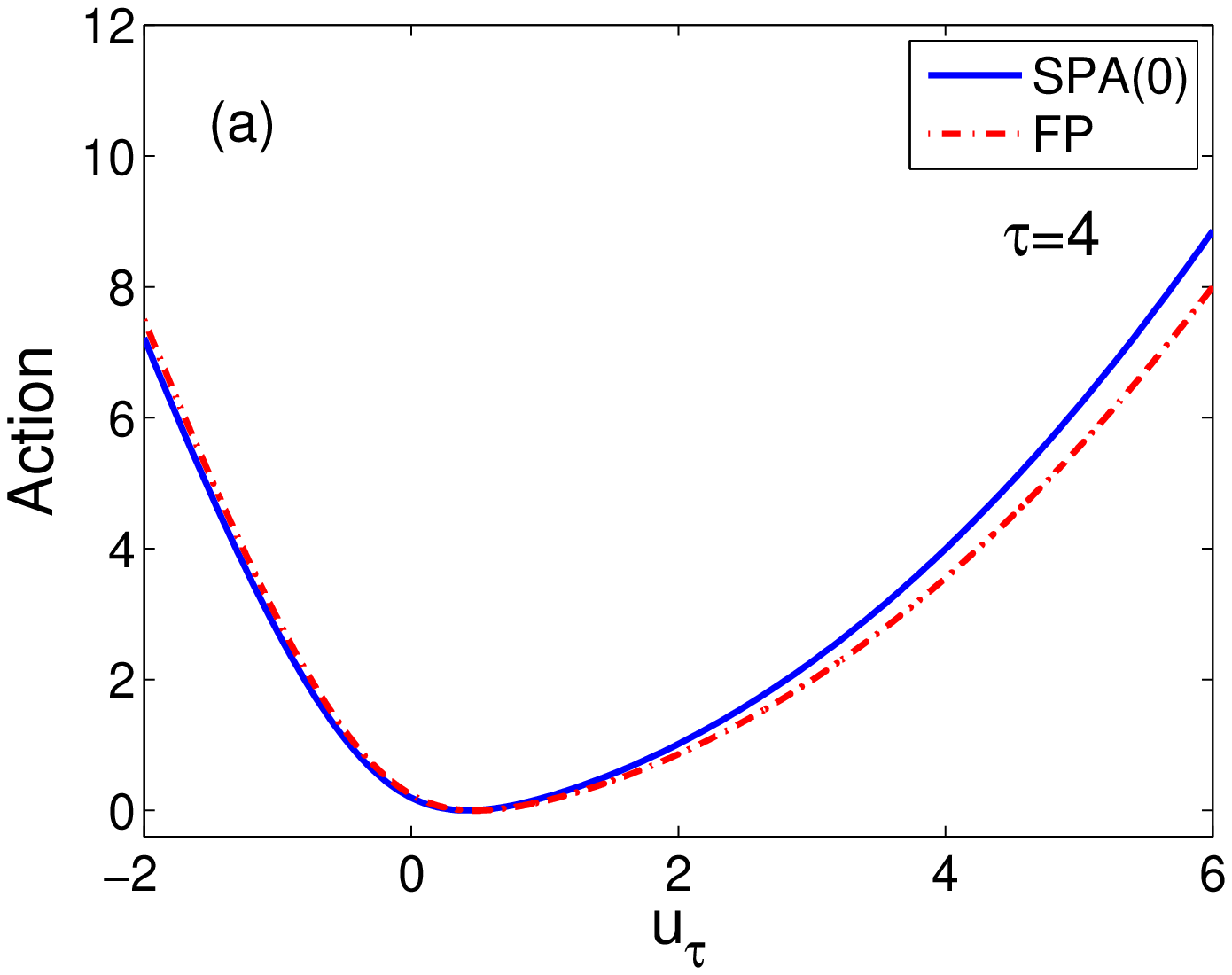}
\includegraphics[scale=0.37]{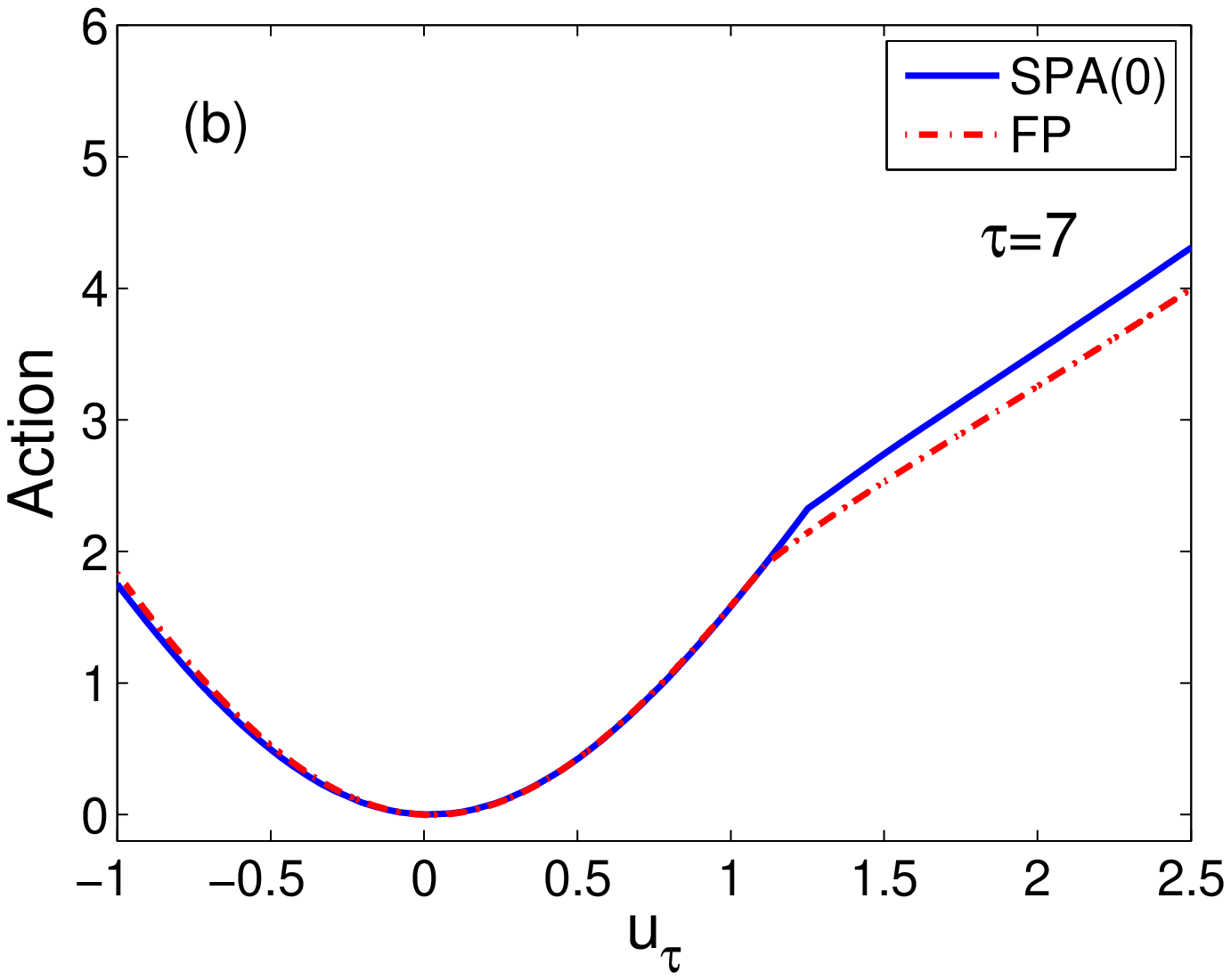}
\includegraphics[scale=0.37]{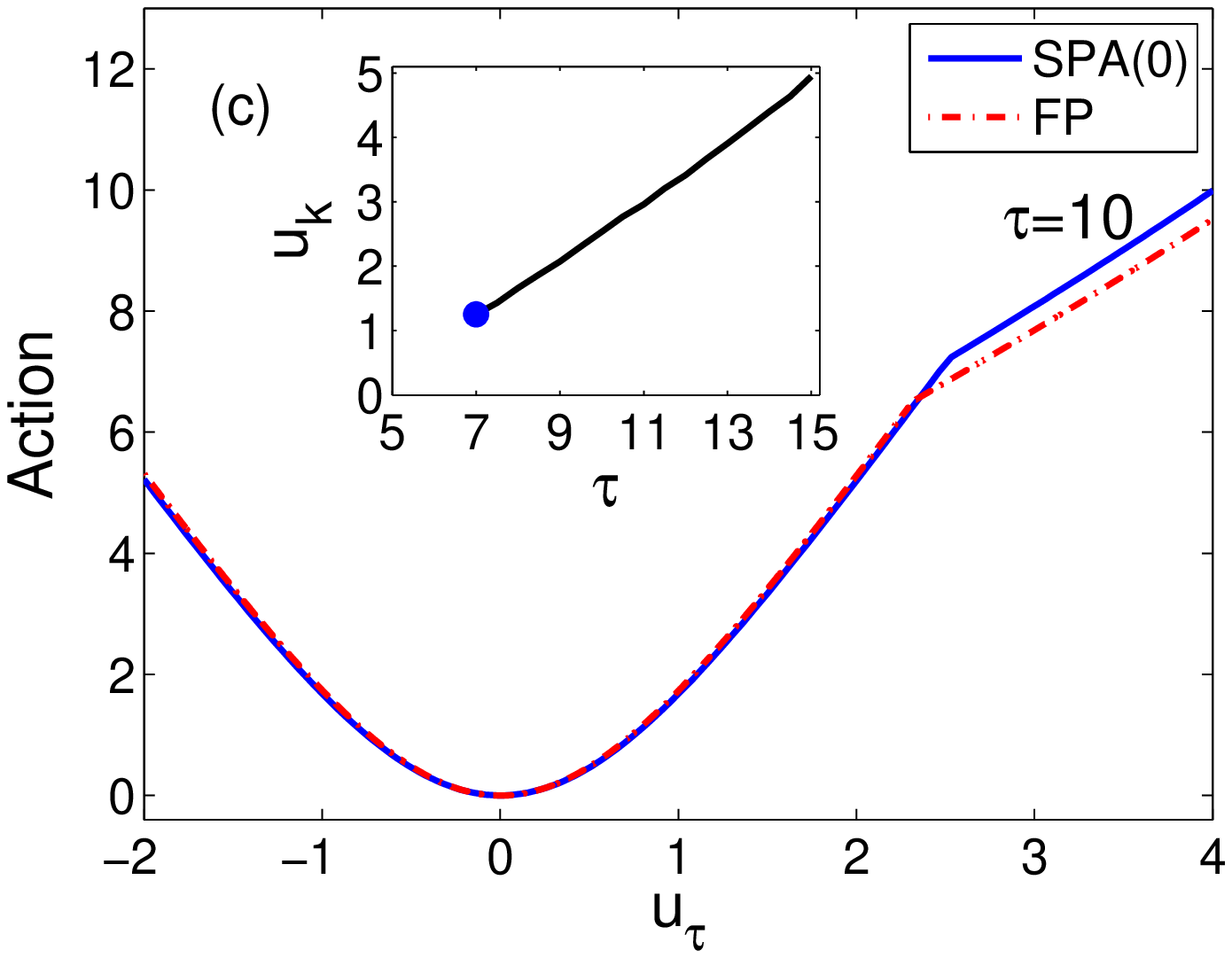}
\end{center}
\caption{(Color online) quasipotential of the regularized model, Eq.~(\ref{EqregularizedLangevin}),
for initial condition $u_0=4$, $ D=0.0001 $ and three different values of time: (a) $\tau=4$, (b) $\tau=7$,
and (c) $\tau=10$. Here FP (dashed line) denotes the potential $-4D\ln p(u_\tau,\tau|u_0,0)$ computed by numerical integration of the Fokker-Planck equation and SPA(0) (solid line) the result of the leading-order SPA, i.e., minimized action (\ref{EqzeroOrderActionTanh}). The inset in (c) shows the position of the kink, $u_k$,  as a function of the time $\tau$, as obtained from the SPA(0). The closed circle shows the time and the position where the kink emerges [see also Fig.~\ref{Figkinksolution}]. }
\label{FigSPA}
\end{figure}

A nonunique solution to the boundary value problem implies that the actual
optimizing path has to be determined from minimizing Eq.~(\ref{EqzeroOrderActionTanh})
among the three possible paths. In our case, Fig.~\ref{FigEndpoints} shows that the optimal path is
either a direct or an indirect path, as for the piecewise-constant SDE.
The intermediate path is always a saddle point of the action, which
implies that the quasipotential
consists of two analytic branches, with a kink appearing when the type of optimal paths changes
[see Figs.~\ref{FigEulerPathEsilon002}(b), \ref{FigSPA}(b), and \ref{FigSPA}(c)].
The kink, which also shows up in the numerical solution of the Fokker-Planck
equation, is an important feature of the quasipotential: It appears at finite time and
then moves to larger $u_{\tau}$ values [see Fig.~\ref{FigSPA}(c)].
This feature, which is related to the convergence of the propagator towards
the stationary distribution, is studied in detail next.

\section{Analytic properties of the action}
\label{sect.analyticalAction}

The previous analysis was mainly based on a numerical solution of the 
boundary value problem
(\ref{EqzeroOrderELTanh}) and the evaluation of the corresponding action 
integral
(\ref{EqzeroOrderActionTanh}) (see \cite{CaKn_PRB81} for a related 
numerical study). In this section, we obtain further insights into the weak-noise limit by
studying some analytical properties of the quasipotential, focusing on
the zeroth-order approximation of the action. The first-order approximation of the action can be
analyzed along similar lines.

\subsection{Action integral}

\begin{figure}
\begin{center}
\includegraphics[scale=0.5]{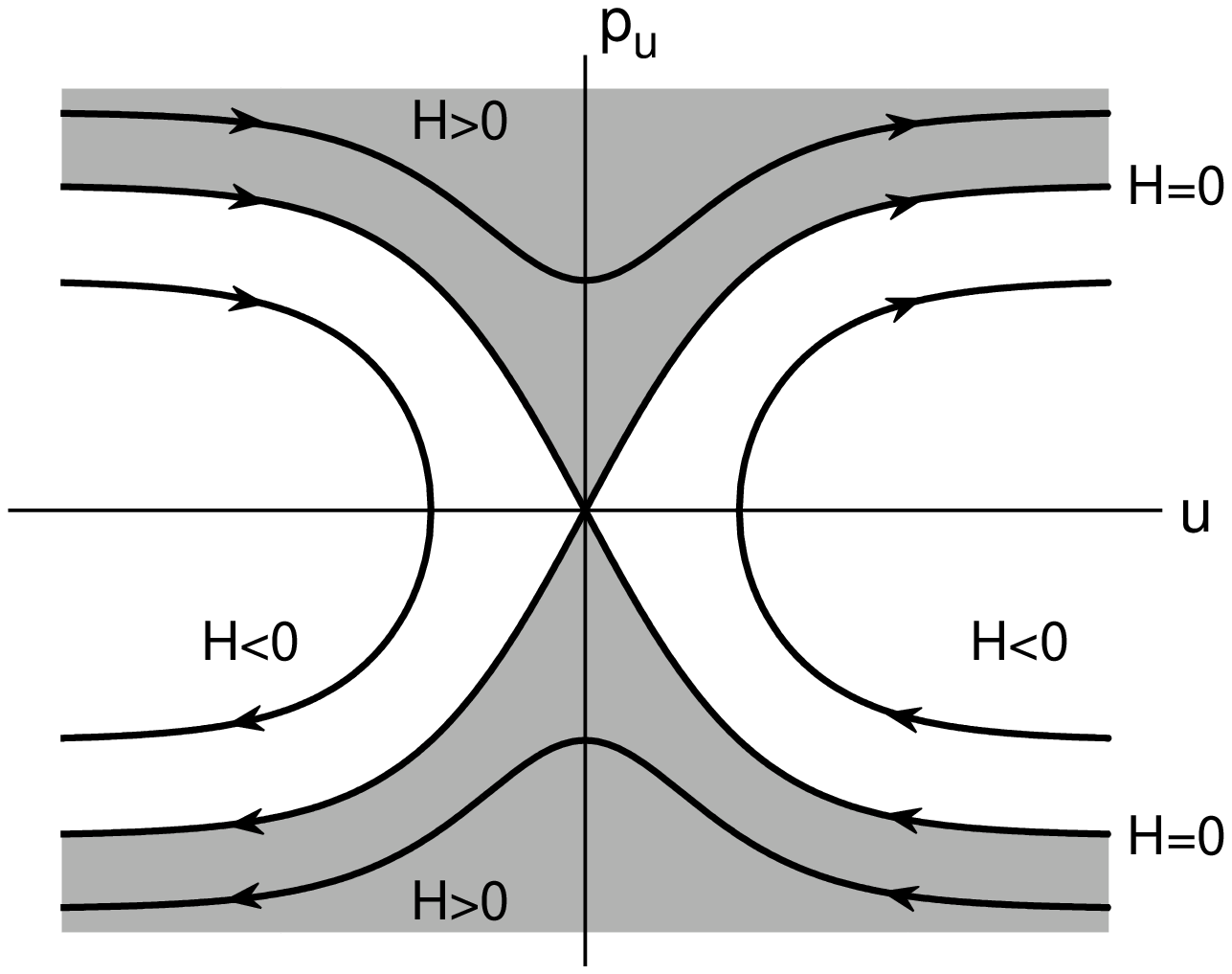}
\end{center}
\caption{Phase portrait of the Hamiltonian system (\ref{EqzeroOrderHamiltonianTanh}) that governs the
boundary value problem (\ref{EqzeroOrderELTanh}). The region of positive energy is shaded.}
\label{Figphase}
\end{figure}

The Euler-Lagrange equations (\ref{EqzeroOrderELTanh})
are equivalent to a conservative system with the Hamiltonian
\begin{equation}\label{EqzeroOrderHamiltonianTanh}
H(p_u,u)=\frac{1}{2}p_u^2-\frac{1}{2}\tanh^2 (u) ,
\end{equation}
where $p_u=\dot{u}$.
The phase portrait of this system, shown in Fig.~\ref{Figphase}, is important to understand the
structure of the boundary value problem. The origin in phase space is a hyperbolic equilibrium point.
Regions of positive and negative energy, respectively, are bounded by the separatrix of this fixed point.
Solutions of the equation of motion (\ref{EqzeroOrderELTanh}) are given by constant energy levels
\begin{equation}\label{EqODE}
\dot{u}=p_u=\pm \sqrt{\tanh^2 (u)+2H},
\end{equation}
where the sign determines whether the path is in the positive or negative momentum
region of phase space. The zeroth-order action (\ref{EqzeroOrderActionTanh}) for a path with
energy $H$, starting at $u_0$ and terminating at $u_\tau$ can then be written as
\begin{equation}\label{EqzeroOrderActionTanhSimplified}
S^{(0)}[u]=S(u_0,u_\tau,\tau,H)
=2\int_{u_0, H}^{u_{\tau}} p_u(s) \mbox{d}u(s)
+2\ln\left(\frac{\cosh (u_{\tau})}{\cosh( u_0)}\right)
-2H\tau
\end{equation}
if we take the expression (\ref{EqzeroOrderHamiltonianTanh}) for the Hamiltonian into account.
The duration $\tau$ of this path from $u_0$ to $u_\tau$ is evaluated as
\begin{equation}\label{Eqtime}
\tau=\int_{u_0, H}^{u_{\tau}}\frac{\mbox{d}u(s)}{p_u(s)},
\end{equation}
and allows us to express the energy in Eq.~(\ref{EqzeroOrderActionTanhSimplified})
in terms of the time. Thus the problem has been reduced to evaluating two integrals
and solving algebraic equations. To take care of the
correct sign in Eq.~(\ref{EqODE}), we have to distinguish three cases,
according to the sign of the terminal point $u_\tau$.

\subsubsection{Case 1: $u_{\tau}<0<u_0$}
\begin{figure}
\begin{center}
\includegraphics[scale=0.5]{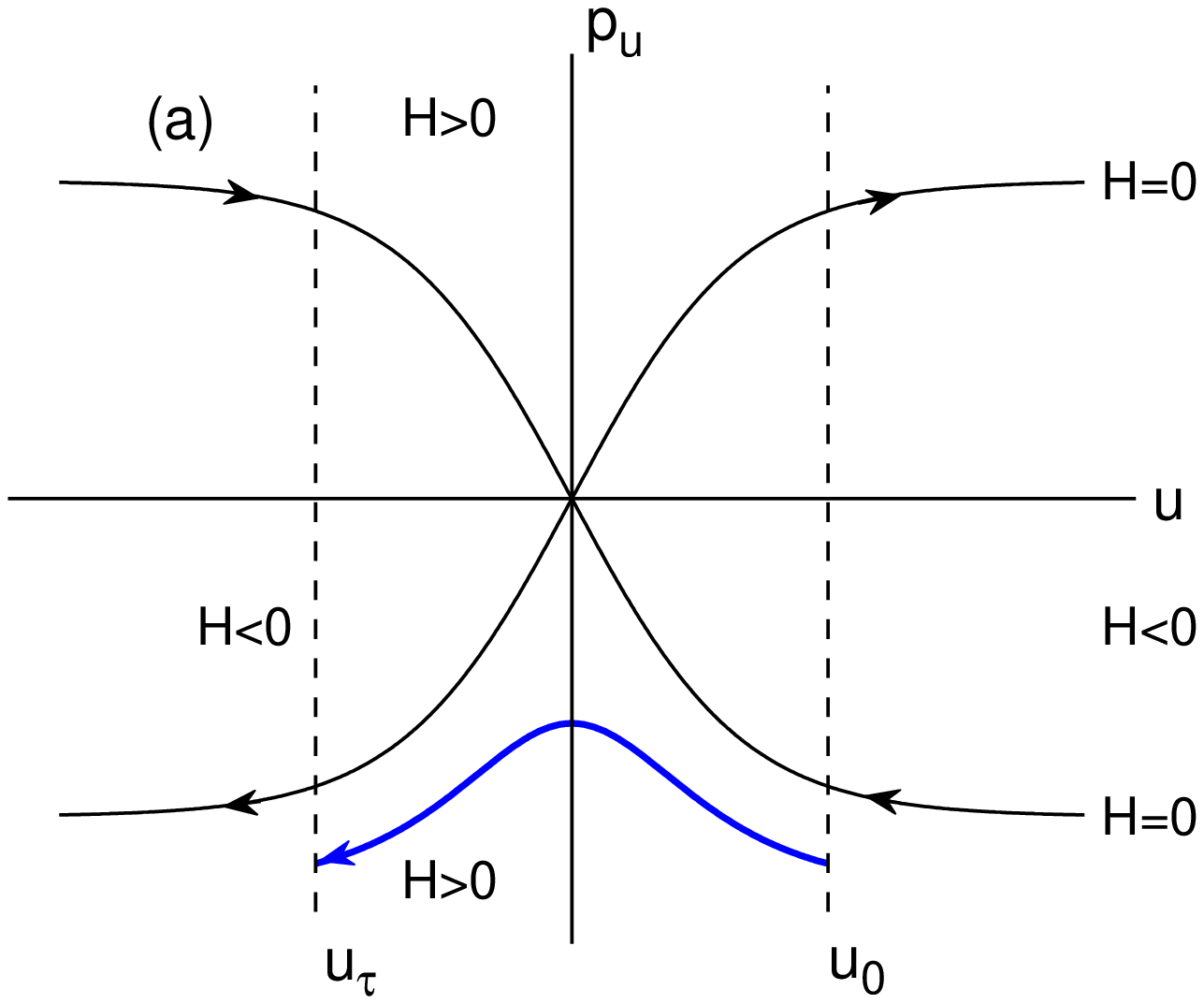}
\includegraphics[scale=0.5]{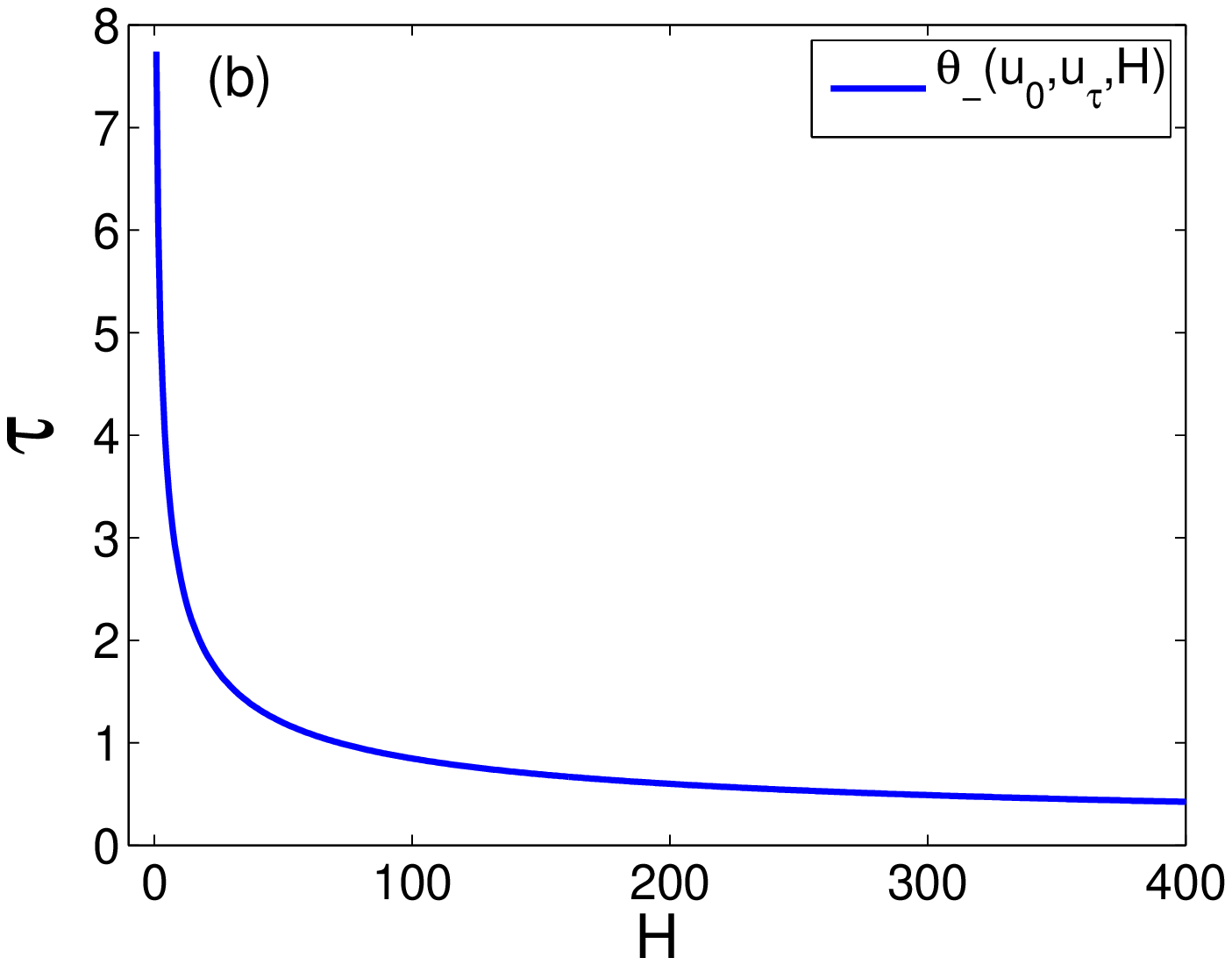}
\end{center}
\caption{(Color online) (a) Phase portrait of the Hamiltonian (\ref{EqzeroOrderHamiltonianTanh})
with an orbit connecting two boundary points
 $ u_{\tau}<0<u_0 $ at a given energy $H>0$.
(b) Energy time relation (\ref{EqtimeCase3}) for $u_0=10$ and $u_\tau=-2$.}\label{Figphase3}
\end{figure}

As can be seen from the phase portrait of Fig.~\ref{Figphase3}(a), there is for given energy
$H>0$ a unique path connecting the two boundary points.
Along this path Eq.~(\ref{EqODE})
holds with a minus sign, and the expression relating energy and time, Eq.~(\ref{Eqtime}),
results in
\begin{equation}\label{EqtimeCase3}
\tau = \int_{u_\tau}^{u_0} \frac{\mbox{d}u}{\sqrt{\tanh^2(u)+2 H}} =
\theta_-(u_0,u_\tau,H),
\end{equation}
where
\begin{equation}\label{db}
\theta_-(u_0,u_\tau,H)=
\frac{1}{\sqrt{1+2 H}} \ln \left(
\frac{\sinh(u_0)+\sqrt{2H/(1+2H) + \sinh^2(u_0)}}{\sinh(u_\tau)+\sqrt{2H/(1+2H) + \sinh^2(u_\tau)}}
\right) .
\end{equation}
It is easy to show that in the range $H>0$ and $u_{\tau}<0<u_0$, Eq.~(\ref{db})
is a monotonically decreasing function of $H$ [see Fig.~\ref{Figphase3}(b)], i.e.,
Eq.~(\ref{EqtimeCase3}) defines the energy $H$ as an
analytic expression of $\tau$ and of the boundary points (see Appendix \ref{AppMonotonicOfTheta}). As for the integral
that enters the action (\ref{EqzeroOrderActionTanhSimplified}) we obtain
\begin{equation}\label{da}
\int_{u_0,H}^{u_\tau} p_u(s) d u(s) = \int_{u_\tau}^{u_0} \sqrt{\tanh^2(u) +2 H} \mbox{d}u
= \int_{u_\tau}^{u_0} \frac{1+2H-1/\cosh^2(u)}{\sqrt{\tanh^2(u) +2 H}} \mbox{d}u 
= (1+2H) \tau- \sigma_-(u_0,u_\tau,H),
\end{equation}
where we have introduced the abbreviation
\begin{equation}\label{daa}
\sigma_-(u_0,u_\tau,H)= \ln\left(\tanh(u_0)+\sqrt{2 H +\tanh^2(u_0)}\right)
-\ln\left(\tanh(u_\tau)+\sqrt{2 H +\tanh^2(u_\tau)}\right).
\end{equation}
Therefore, the entire action (\ref{EqzeroOrderActionTanhSimplified}) finally reads
\begin{equation}\label{EqzeroOrderActionTanhCase3}
S(u_0,u_\tau,\tau,H)= 2(1+H)\tau+2\ln\left(\frac{\cosh (u_{\tau})}{\cosh (u_0)}\right)
-2 \sigma_-(u_0,u_\tau,H).
\end{equation}
Since the boundary value problem has a unique solution, the inverse of the energy-time relation
(\ref{EqtimeCase3}) is single valued, so that the expression for the action defines an analytic expression of time $\tau$ and
of the boundary points in the region $u_\tau<0<u_0$ (see as well Fig.~\ref{FigSPA}).

\begin{figure}
\begin{center}
\includegraphics[scale=0.37]{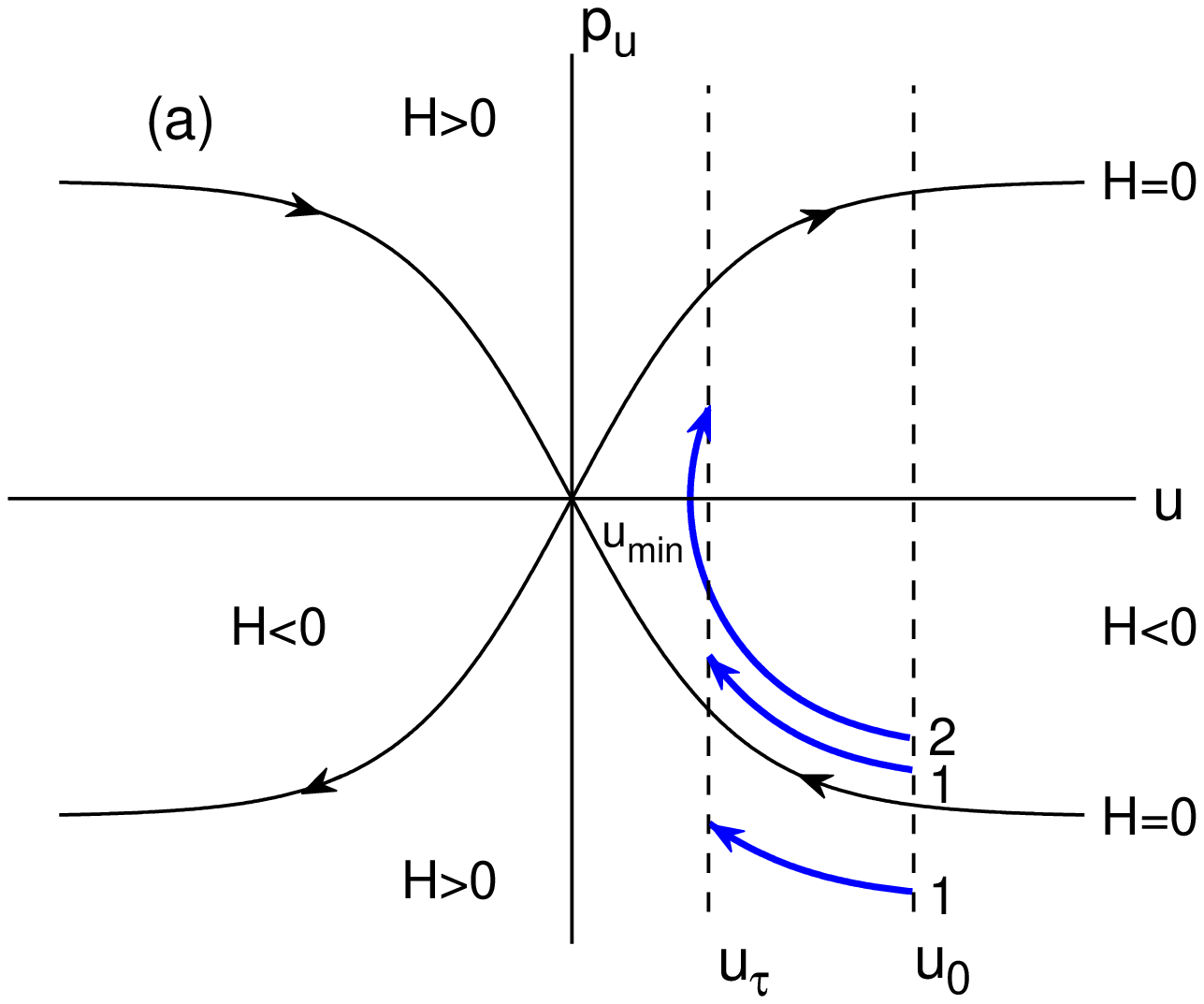}
\includegraphics[scale=0.37]{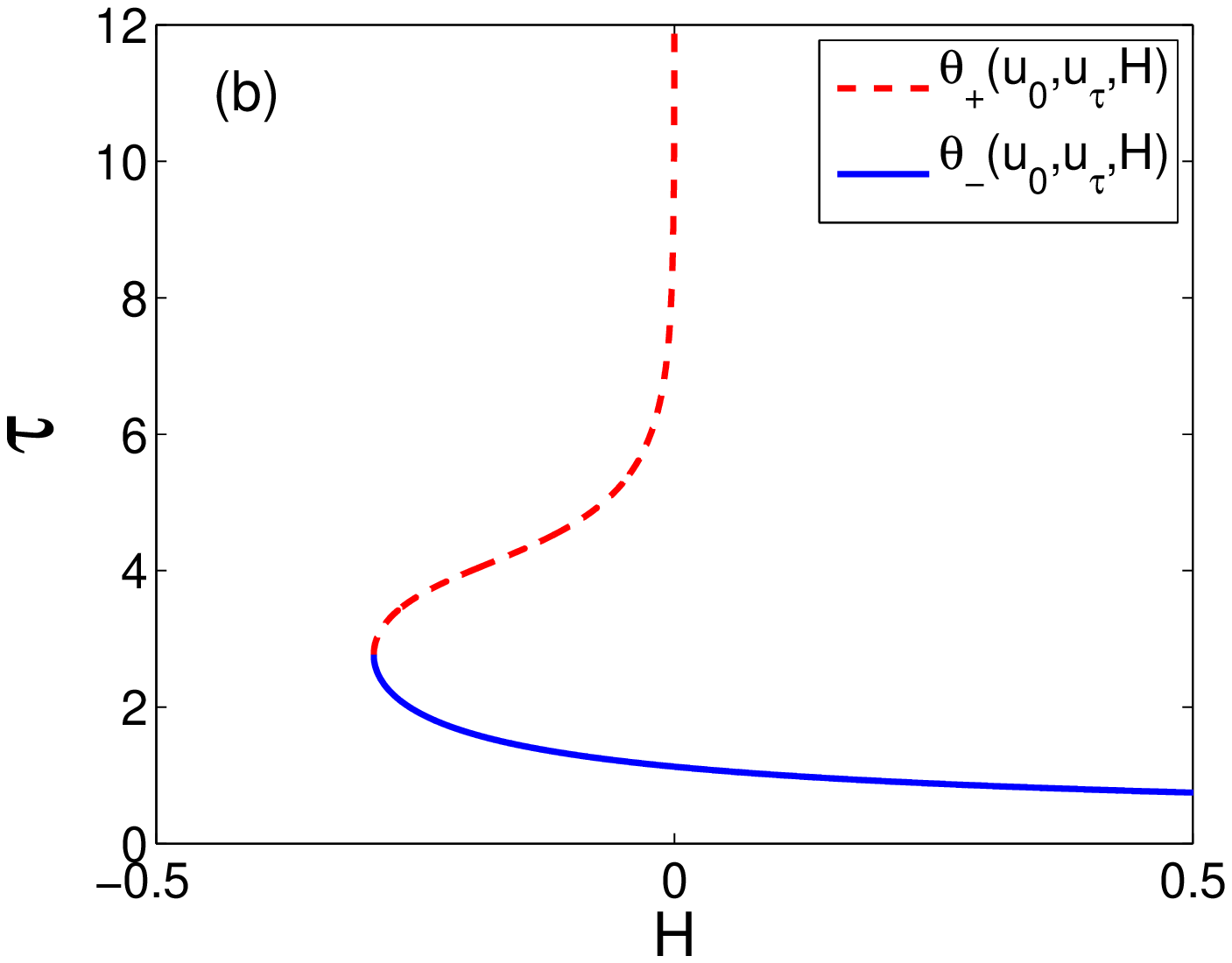}
\includegraphics[scale=0.37]{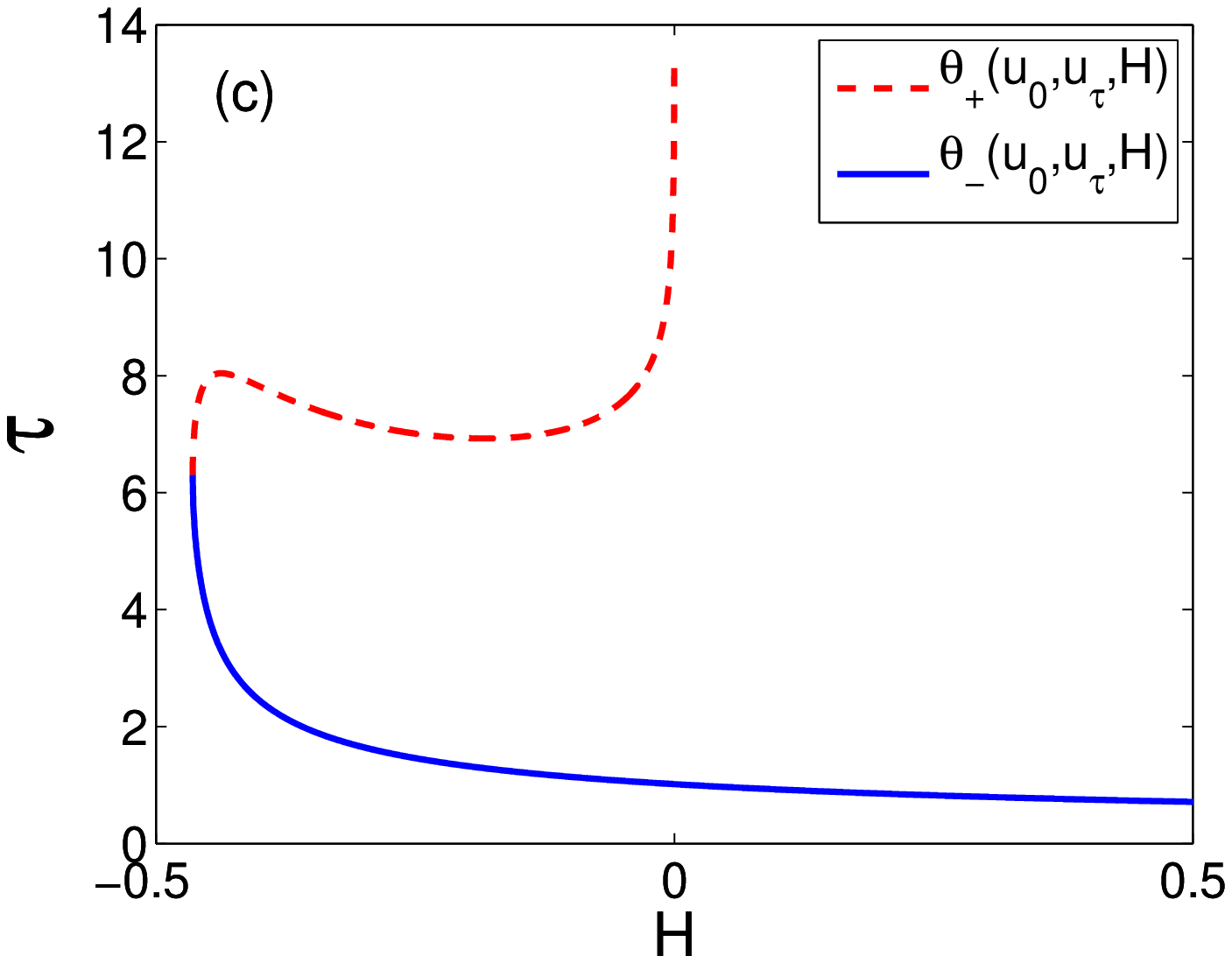}
\end{center}
\caption{(Color online) (a) Phase portrait of the Hamiltonian (\ref{EqzeroOrderHamiltonianTanh})
with direct (1) and indirect (2) paths
connecting boundary points $ 0<u_{\tau}<u_0 $.
(b) Energy time relation (\ref{EqtimeCase3}) (solid line) and (\ref{dc}) (dashed line) for
small values of $u_0$ ($u_0=2$ and $u_\tau=1$) with a unique inverse.
(c) Energy time relations (\ref{EqtimeCase3}) (solid line) and (\ref{dc}) (dashed line) for
larger  values of $u_0$ ($u_0=3$ and $u_\tau=2$) with a multivalued inverse in a finite time
interval. }\label{Figphase2}
\end{figure}

\subsubsection{Case 2: $0<u_{\tau}<u_0$}\label{case2}
For boundary points $0<u_{\tau}<u_0$ the phase portrait in Fig.~\ref{Figphase2}(a)
and Eq.~(\ref{EqzeroOrderHamiltonianTanh}) show that the
connecting path has energy $H>H_{min}=-\tanh^2(u_\tau)/2$. On the one hand, for energies $H>H_{min}$, there exists a direct path whose duration
is given by Eq.~(\ref{EqtimeCase3}) and whose action is determined by Eq.~(\ref{EqzeroOrderActionTanhCase3}).
On the other hand, for $H_{min}<H<0$, there exists an indirect path with a turning point at
$u_{min}=\mbox{artanh}(\sqrt{-2H})$. The duration of this path is obtained from Eq.~(\ref{Eqtime}) by
splitting the path into two parts and using appropriate signs in Eq.~(\ref{EqODE}):
\begin{eqnarray}\label{dc}
\tau = \int_{u_0}^{u_{min}} \frac{\mbox{d} u}{-\sqrt{ \tanh^2 u+2H }} +
\int_{u_{min}}^{u_{\tau}}\frac{\mathrm{d}u}{ \sqrt{ \tanh^2 u+2H } }
= \theta_+(u_0,u_\tau,H),
\end{eqnarray}
where
\begin{eqnarray}\label{dd}
\theta_+(u_0,u_\tau,H) &=&
\frac{\ln \left(\sinh(u_0)+\sqrt{2H/(1+2H) + \sinh^2(u_0)}\right) }{\sqrt{1+2 H}}+
\frac{\ln\left(\sinh(u_\tau)+\sqrt{2H/(1+2H) + \sinh^2(u_\tau)}\right)}{\sqrt{1+2 H}} \nonumber \\
& & +
\frac{\ln\left((1+2H)/(-2H)\right) }{\sqrt{1+2 H}} .
\end{eqnarray}
Similarly, the integral appearing in the action (\ref{EqzeroOrderActionTanhSimplified})
can be evaluated [see Eq.~(\ref{da})]
\begin{eqnarray}\label{de}
\int_{u_0,H}^{u_\tau} p_u(s) \mbox{d}u(s)  =  -\int_{u_0}^{u_{min}} \sqrt{2H +\tanh^2(u)} \mbox{d}u
+ \int_{u_{min}}^{u_\tau} \sqrt{2H +\tanh^2(u)} \mbox{d}u
 =  (1+2 H)\tau - \sigma_+(u_0,u_\tau,H),
\end{eqnarray}
where we have introduced the abbreviation
\begin{eqnarray}\label{dea}
\sigma_+(u_0,u_\tau,H) = \ln\left(\tanh(u_0)+\sqrt{2 H +\tanh^2(u_0)}\right)+
\ln\left(\tanh(u_\tau)+\sqrt{2 H +\tanh^2(u_\tau)}\right)
  -\ln(-2H).
\end{eqnarray}
Thus, for the action (\ref{EqzeroOrderActionTanhSimplified}) of this path we obtain
\begin{equation}\label{df}
S(u_0,u_\tau,\tau,H) = 2(1+H)\tau +2\ln\left(\frac{\cosh (u_{\tau})}{\cosh (u_0)}\right)
-2 \sigma_+(u_0,u_\tau,H).
\end{equation}

\begin{figure}
\begin{center}
\includegraphics[scale=0.5]{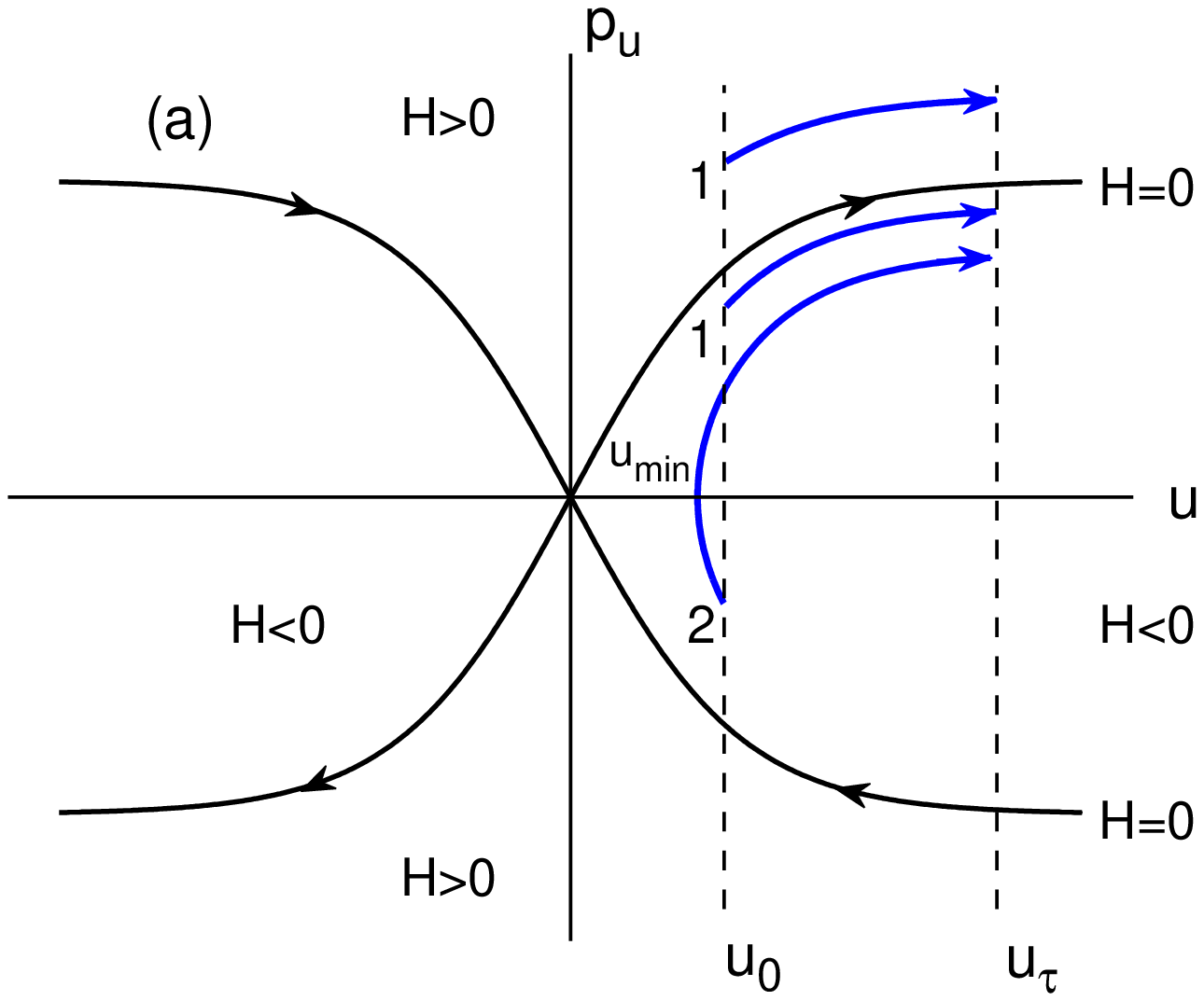}
\includegraphics[scale=0.5]{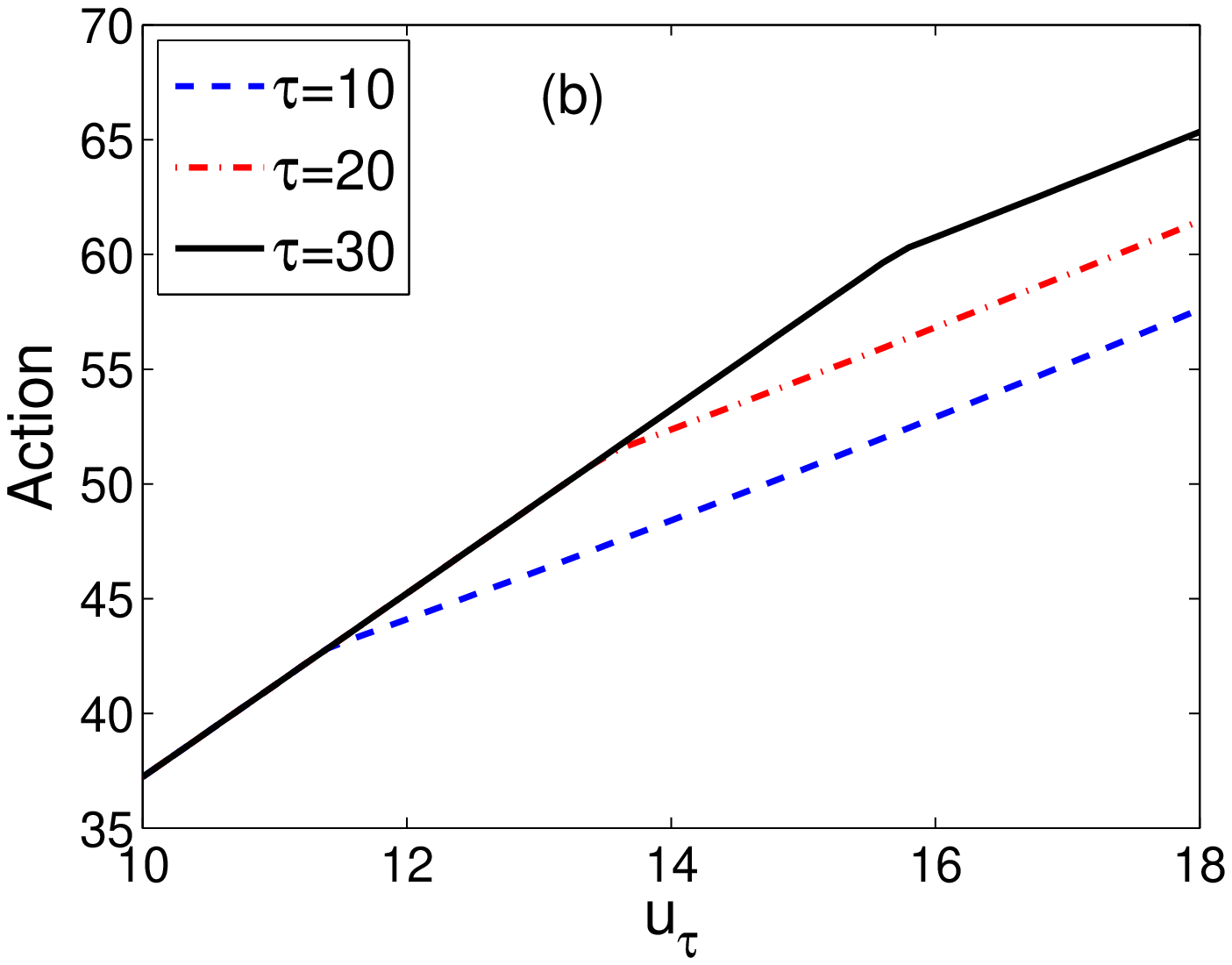}
\end{center}
\caption{(Color online) (a) Phase portrait of the Hamiltonian (\ref{EqzeroOrderHamiltonianTanh})
with direct (1) and indirect (2) orbits connecting boundary points
$ 0<u_0<u_{\tau} $. (b) Minimal action (\ref{EqzeroOrderActionTanhSimplified}) as a function of $u_\tau$  in the interval
$u_\tau>u_0=10$ for different values of the time $\tau$, showing a kink moving to the right.}
\label{Figphase1}
\end{figure}

To find the minimizing path, we have to take a closer look at the energy-time relation.
This relation consists of two branches [see Fig.~\ref{Figphase2}(b) and \ref{Figphase2}(c)].
The branch determined by Eq.~(\ref{EqtimeCase3})
is defined for all energies $H_{min}<H$ and is a monotonically decreasing function for $0<u_\tau<u_0$
(see Appendix \ref{AppMonotonicOfTheta}). The second
branch, defined for negative energies $H_{min}<H<0$ only, is given by Eq.~(\ref{dc}).
This branch is monotonically increasing for small values of $ u_0 $ and develops an
inflection point if the boundary points exceed a critical value, which implies that, for small values of $ u_0 $, the relation determines the energy in terms of time and boundary points uniquely. This uniqueness is expected, as the analysis of the regularized model reduces, in the limit of small value of $u_0$, to that of the
Ornstein-Uhlenbeck process. However, if the initial condition is beyond the linear regime, the energy
time relation has a multivalued inverse within a finite-time interval [see Fig.~\ref{Figphase2}(c)].
The value of the action entering in the propagator is then the minimum among the three possible inverse values.
The data show that for fixed values of $u_0$ and $u_\tau$ the minimum is
given by the low energy solution, i.e., by the direct path, up to a critical value of $\tau$, while the minimum
for larger values of $\tau$ is given by the high-energy solution, i.e., the indirect path.
The intermediate path, i.e., the solution with energy in between, is always of saddle type.
As a consequence, the action, considered as a function of $u_\tau$ switches the analytical branch and
develops a kink [see Figs.~\ref{FigSPA}(b), \ref{FigSPA}(c), and \ref{Figphase1}(b)].

\subsubsection{Case 3: $0 < u_0<u_{\tau}$}
As in the previous case there are two types of paths to consider: In the energy range $H>H_{min}=-\tanh^2(u_0)/2$,
there exists a direct path connecting the two boundary points, while for negative energies $H_{min}<H<0$,
there exist in addition indirect paths with a turning point at $u_{min}=\mbox{artanh}(\sqrt{-2H})$. This is 
similar to case 2 and the relevant integrals can be dealt with by applying the symmetry of the phase portrait
[see Fig.~\ref{Figphase1}(a)]. To find the duration of the direct path (\ref{Eqtime}) we observe that just by swapping the boundary points
and reversing the direction of the path we obtain a corresponding mirror orbit with the same duration,
which we have dealt with in the previous paragraph. Thus Eq.~(\ref{EqtimeCase3}) tells us that
\begin{equation}\label{EqtimeCase1Path1}
\tau=\int_{u_0}^{u_{\tau}}\frac{\mbox{d}u}{ \sqrt{ \tanh^2 u+2H } }=
\theta_-(u_\tau,u_0,H) .
\end{equation}
Since the integral of the action has the same symmetry, we obtain the result stated in Eq.~(\ref{da}) by swapping $u_0$ and $u_\tau$ on the right-hand side. Thus the action
(\ref{EqzeroOrderActionTanhSimplified}) finally reads
\begin{equation}\label{EqzeroOrderActionTanhCase1Path1}
S(u_0,u_\tau,\tau,H) = 2(1+H)\tau+2\ln\left(\frac{\cosh (u_{\tau})}{\cosh (u_0)}\right)
-2 \sigma_-(u_\tau,u_0,H).
\end{equation}
The action itself does not share the aforementioned symmetry because of the additional terms appearing
in Eq.~(\ref{EqzeroOrderActionTanhSimplified}).

For the indirect path the same reasoning applies. The duration of the path, given by Eq.~(\ref{dc}), applies since the
right-hand side is symmetric in $u_0$ and $u_\tau$ and swapping both arguments does not have any effect.
In fact, the same argument is valid for the integral that enters the action. Therefore, Eq.~(\ref{de})
holds since the right-hand side is a symmetric expression in the boundary points and the corresponding action
is indeed given by Eq.~(\ref{df}).

Overall, we see that the energy-time relationship is at the heart of understanding
the structure of the optimal paths, the minimized action, and finally the propagator of
the SDE. As in the previous case, the relation consists of two branches
[see Figs.~\ref{Figphase2}(b) and \ref{Figphase2}(c)].
The lower branch is given by the direct path (\ref{EqtimeCase1Path1}),  which is a monotonically decreasing
function on its domain $H>H_{min}$. The second branch for negative energies $H_{min}<H<0$
is determined by Eq.~(\ref{dc}) and thus largely the discussion of the previous section applies.
If the initial condition is small, so that no cubic singularity appears
for $0<u_\tau<u_0$ (case 2), then such a singularity will finally develop for sufficiently large
value of $u_\tau$ resulting in a non analytic minimized action. If, on the contrary, $u_0$ is
so large that the occurrence of the kink has already happened in the domain $0<u_\tau<u_0$,
then the kink just propagates to larger values of $u_\tau$
[see Fig.~\ref{Figphase1}(b)].

\subsection{Asymptotics of the action}

The analytic expressions derived in the previous section allow us to study in
some detail the properties of the propagator shown, e.g., in Fig.~\ref{FigSPA}. 
Of particular relevance is the approach to the stationary state, the tail behavior of the distribution,
and the emergence and dynamics of nonanalyticities of the quasipotential.

\subsubsection{Approach to equilibrium}

To study the convergence of the propagator towards the stationary distribution, we
consider a fixed value of $u_0$ and  $u_\tau$ and take the asymptotic limit $\tau\rightarrow \infty$.
It is obvious from the phase portrait or from the energy-time relation [see
Figs.~\ref{Figphase3}(b) and \ref{Figphase2}(b)] that the asymptotic limit implies $ H\rightarrow 0 $.
Hence the energy-time relation is determined by Eq.~(\ref{EqtimeCase3}) or (\ref{dc}), depending on the sign of
$u_\tau$. In both cases, a straightforward expansion yields
\begin{equation}\label{ha}
\tau= \ln(2 \sinh(u_0))+\ln(-\sinh(u_\tau)/H)+O(H \ln|H|) \, .
\end{equation}
For the action, either Eq.~(\ref{EqzeroOrderActionTanhCase3}) or
(\ref{df}) applies, depending on the sign of $u_\tau$, yielding the expansion
\begin{eqnarray}\label{hb}
S(u_0,u_\tau,\tau,H) = 2(1+H)\tau+2\ln\left(\frac{\cosh (u_{\tau})}{\cosh (u_0)}\right)
 -
2 \left(\ln\left(2\tanh(u_0)\right) + \ln \left(-\tanh(u_\tau)/H\right)\right) + O(H).
\end{eqnarray}
By solving Eq.~(\ref{ha}) to leading order for $H$, Eq.~(\ref{hb}) yields, as expected, the potential of
the stationary distribution [see also Eq.~(\ref{EqzeroOrderPropagator})]
\begin{equation}\label{hc}
S(u_0,u_\tau,\tau,H) = 4\ln \cosh(u_\tau) + O(\tau \exp(-\tau)). 
\end{equation}
The full action as well as the propagator at finite time depends on the initial condition  $ u_0 $ and hence
does not have a symmetry with respect to $ u_\tau $. However, the stationary action and the corresponding stationary distribution is independent of $ u_0 $ [see Eq.~(\ref{hc})] in the limit $\tau \rightarrow \infty$. In that limit
the symmetry under the change $ u_\tau\rightarrow -u_\tau $ is restored [see as well 
Eq.~(\ref{EqStationarySolutionDryFriction})]. Above all, the saddle-point expansion preserves all the symmetries
of the underlying equations of motion. As for the transient properties it is easy, though more tedious, to compute 
the subleading term in Eq.~(\ref{hc}), which then quantitatively describes the relaxation process.

\subsubsection{Tail behavior of the propagator}

To investigate the action for large values of arguments, we consider
the asymptotic limit $|u_\tau| \rightarrow \infty$ for fixed values of the time $\tau$ and initial condition $u_0$, which implies, from the phase portraits of
Figs.~\ref{Figphase3}(a) and \ref{Figphase1}(a), $ H\rightarrow +\infty $.
The energy-time relation is determined by Eqs.~(\ref{EqtimeCase3}) and (\ref{EqtimeCase1Path1}),
depending on the sign of $u_\tau$, and a direct expansion results in the expression
\begin{equation}\label{hd}
\tau = \frac{1}{\sqrt{1+2 H}}\left(|u_\tau-u_0| + O(H^{-1}) \right),
\end{equation}
which is valid irrespective of the sign of $u_\tau$. This expression is easily inverted to obtain
\begin{equation}\label{he}
1+2H = (u_\tau-u_0)^2/\tau^2 + O(u_\tau^{-1}).
\end{equation}
For the action Eqs.~(\ref{EqzeroOrderActionTanhCase3}) or (\ref{EqzeroOrderActionTanhCase1Path1}) apply,
depending on the sign of $u_\tau$.
Expansion in terms of $H$ yields $\sigma_-$ to be of order $O(H^{-1})$ so that
\begin{equation}\label{hf}
S(u_0,u_\tau,\tau,H) = 2(1+H) \tau +2\ln\left(\frac{\cosh (u_{\tau})}{\cosh (u_0)}\right) + O(H^{-1}).
\end{equation}
Hence, using Eq.~(\ref{he}) we end up with
\begin{equation}\label{hg}
S(u_0,u_\tau,\tau,H) = \frac{(u_{\tau}-u_0)^2}{\tau}+\tau+2\ln\left(\frac{\cosh(u_{\tau})}{\cosh(u_0)}\right)
+ O(u_\tau^{-1}).
\end{equation}
To leading order, the tails of the propagator have a Gaussian shape, as expected, while the subleading terms
indicate a transition from the Gaussian shape to the exponential shape of the stationary distribution.

\subsubsection{Emergence and propagation of the kink}

\begin{figure}
\begin{center}
\includegraphics[scale=0.5]{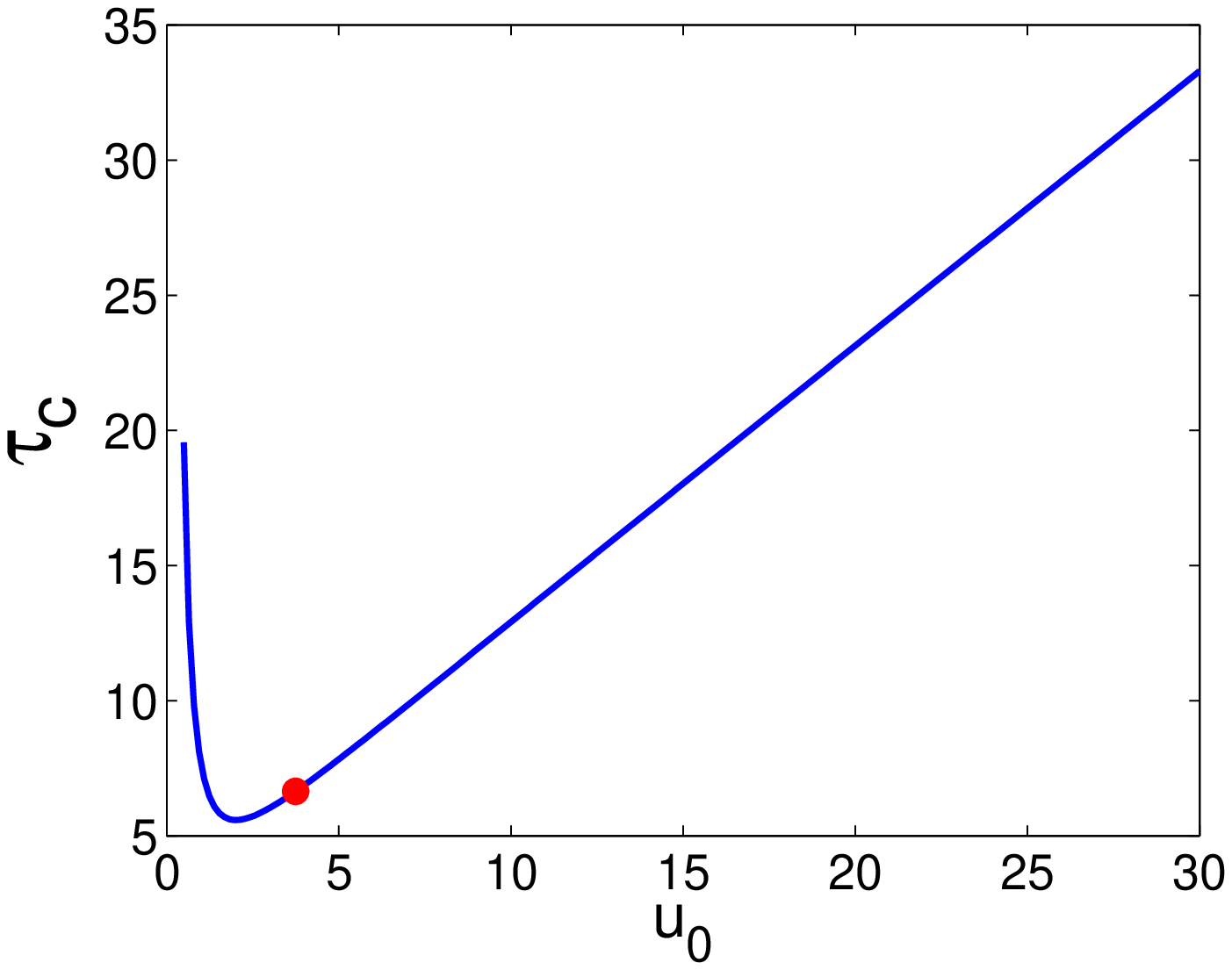}
\includegraphics[scale=0.5]{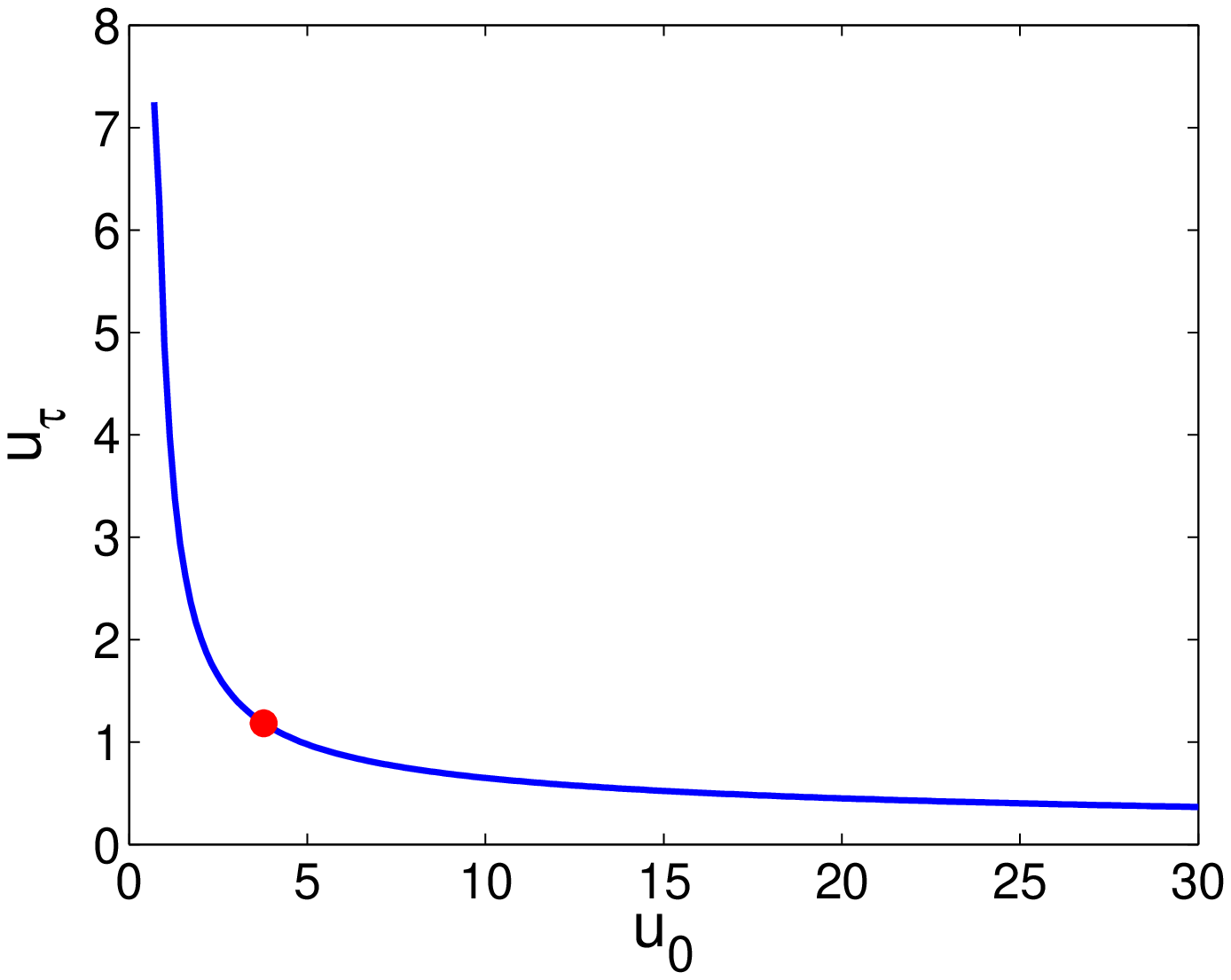}
\end{center}
\caption{(Color online) Time $\tau_c$ and position $u_\tau$ for the emergence of the kink in the minimized action
as a function of the initial condition $u_0$
[see Eqs.~(\ref{EqKinkTime})]. The closed circle indicates the starting point of the kink for $u_0=4 $, as in Fig.~\ref{FigSPA}(c). }
\label{Figkinksolution}
\end{figure}

As mentioned before, a kink in the minimized action appears when the energy-time relation has multivalued
inverse. Since $ \theta_-(u_0,u_{\tau},\tau,H) $ is a monotonically decreasing function of $ H $, the kink appears when
$ \theta_+(u_0,u_{\tau},\tau,H) $ develops a crossover from a monotonically increasing shape into a cubic
shape [see Figs.~\ref{Figphase2}(b) and \ref{Figphase2}(c)]. Hence the time $ \tau_c $ when the kink appears and its position $u_\tau$ as a function of initial position $u_0$ are determined by
\begin{equation}\label{EqKinkTime}
\tau=\theta_+(u_0,u_{\tau},H),\qquad \frac{\partial \theta_+}{\partial H}=0, \qquad \frac{\partial^2 \theta_+}{\partial H^2}=0.
\end{equation}
Figure~\ref{Figkinksolution} shows the solution of these equations. For small values of $u_0$, i.e., in the
limiting case of the Ornstein-Uhlenbeck process, both the time and the position for the emergence of the
kink become large. For large values of $u_0$, i.e., when the drift is almost constant, the time
becomes large again but the kink appears close to the origin. At an intermediate scale of the size
of the regularization length the time for the occurrence of the kink becomes minimal and
the kink appears close to the initial condition.

\begin{figure}
\begin{center}
\includegraphics[scale=0.5]{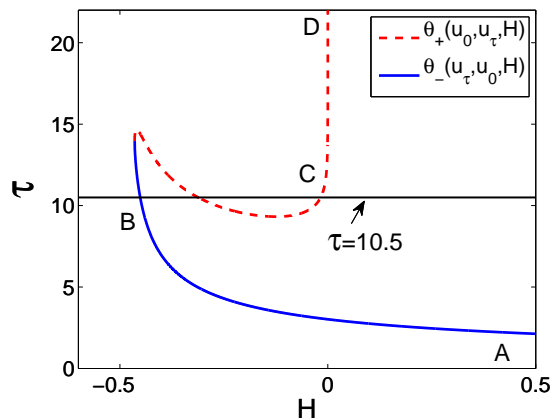}
\end{center}
\caption{(Color online) Energy time relation (\ref{EqtimeCase1Path1}) and (\ref{dc})  for $ u_0=2 $ and $ u_\tau =5 $. The action takes its minimum on the lower branch $ AB $ if $ \tau<10.5 $ and on the upper branch $ CD $ if $ \tau>10.5 $. }
\label{FigThetaLargeUtau}
\end{figure}

After its appearance the kink starts to move with positive velocity [see Figs.~\ref{Figphase1}(b) and \ref{FigSPA}(c)].
We can quantify such a dynamical feature in terms of an asymptotic expansion for large $\tau$ and
large $u_\tau$, keeping the ratio $c=u_\tau/\tau$ of $O(1)$. The key to the analysis is
the shape of the energy-time relation in such a limit (see Fig.~\ref{FigThetaLargeUtau}).

The relation consists of the monotonic
lower branch (\ref{EqtimeCase1Path1}) determined by $\theta_-$ and a cubic shaped upper branch (\ref{dc}) determined by $\theta_+$. Depending on the value of $\tau$ the action takes its minimum either on the
lower branch at an energy $H_-$ that stays of order $O(1)$ or on the upper branch at an
energy $H_+$ that tends to zero in the asymptotic limit considered here. Hence  Eqs.~(\ref{EqtimeCase1Path1}) and
(\ref{dc}) yield
\begin{eqnarray}\label{ia}
\tau &=& \theta_-(u_\tau,u_0,H_-) = \frac{u_\tau}{\sqrt{1+2H_-}} + O(1), \label{iaa} \\
\tau &=& \theta_+(u_\tau,u_0,H_+) = u_\tau+\ln(2 \sinh(u_0)) - \ln(-2 H_+)
+O(\exp(-2 u_\tau), H_+).
\label{iab}
\end{eqnarray}
Both expressions are easily inverted to obtain
\begin{equation}\label{ib}
1+2 H_- = (u_\tau/\tau)^2 +O(\tau^{-1})
\end{equation}
and
\begin{equation}\label{ic}
-2 H_+ = 2 \sinh(u_0) \exp(u_\tau-\tau)
\left(1+O(\exp(-2 u_\tau), \exp(-\tau(1-c))) \right),
\quad (c=u_\tau/\tau<1),
\end{equation}
where for Eq.~(\ref{ic}) we have to require that $c=u_\tau/\tau<1$, otherwise the matching condition
for the action cannot be met since the branch $\theta_+$ has a minimum at finite time
(see Fig.~\ref{FigThetaLargeUtau}).

As for the value of the action along the lower branch,  Eq.~(\ref{EqzeroOrderActionTanhCase1Path1})
yields, if we observe that $\sigma_-$ is of order $O(1)$,
\begin{equation}\label{id}
S_-=2(1+H-) \tau + 2 u_\tau +O(1) = \tau \left( \frac{u_\tau}{\tau} +1 \right)^2 + O(1),
\end{equation}
where we have used Eq.~(\ref{ib}) as well. For the action along the upper branch (\ref{df}) we
have to take into account that $\sigma_+$ can be expressed as $-\ln(-2H_+)+O(1)$ so that
\begin{eqnarray}\label{ie}
S_+ &=& 2 \tau + 2 u_\tau + 2 \ln(-2 H_+) + O(1) = 4 u_\tau+ O(1),
\end{eqnarray}
where we have employed Eq.~(\ref{ic}). The kink's position is determined by equating the expressions
(\ref{id}) and (\ref{ie}), resulting in $u_\tau=\tau$, so that the kink moves with unit speed.
Finally, the minimized action to lowest order reads
\begin{equation}\label{hk}
S=\left\{ \begin{array}{lcl}
4 u_\tau & \mbox{ for } u_\tau<\tau \\
\tau (u_\tau/\tau +1)^2 & \mbox{ for } u_\tau>\tau,
\end{array} \right.
\end{equation}
which indeed describes the numerical findings shown in Fig.~\ref{FigSPA}.

\section{Conclusion}
\label{sect.conclusion}

We have studied in this paper a piecewise-constant SDE in order to understand to what extent
saddle-point approximations of the path integral representation of the propagator, which is
the basis of weak-noise approximations, can be carried out. The advantage of the SDE that we
have considered is that its propagator is known exactly, so that any saddle-point approximations
can be compared to the exact result.

For this model, we have seen that saddle-point approximations are able to reproduce some
features of the propagator, such as its tail behavior and its convergence towards the stationary
distribution, but not the bimodality of the exact propagator appearing at intermediate times. We have
also seen that the lowest-order approximation yields the correct large-deviation approximation
of the propagator, implying that the low-noise large-deviation theory of Freidlin-Wentzell can be applied to this singular SDE.
However, the construction of higher-order corrections to this
approximation is plagued by a fundamental singularity of the Jacobian term of the path integral.

To remove this singularity, we have regularized the discontinuity of the SDE with a smooth nonlinear
drift involving a small parameter controlling the limit to the piecewise smooth drift. The price paid
for introducing this regularization is
that the two limits, small diffusion and small regularization, do not commute and that the propagator
of the regularized SDE is no longer known exactly. Nevertheless, we have shown with simulation results
that the low-noise limit of the regularized SDE captures the
main features of the piecewise smooth SDE. In particular, the orbit or optimal path structure of the regularized
SDE in terms of direct, indirect and intermediate paths is similar to the
optimal path structure inferred heuristically for the piecewise smooth SDE. In addition,
the analysis of the regularized SDE justifies the heuristic
principles that we have defined and used to perform the saddle-point approximation of the piecewise smooth SDE.
For the regularized SDE considered here,
we have finally been able to study the quasipotential associated with the propagator in a largely analytical way.
This is one of the few models for which such results can be obtained. In a future study, we plan to consider higher-dimensional systems, such as models in which inertia is present.

\begin{acknowledgments}
Y.C. was supported by the Chinese Scholarship Council, the Hunan Provincial Innovation
Foundation for Postgraduates (Grant No.~CX2011B011), and NUDT's Innovation Foundation (Grant No.~B110205). W.J. gratefully acknowledges support from EPSRC through Grant No.~EP/H04812X/1 and No.~SFB910 and the kind hospitality by Eckehard Sch\"oll and his group during a stay at TU Berlin.
\end{acknowledgments}


\appendix

\section{Weak-noise approximation of path integrals}
\label{App.pathIntegral}

Path integral formulations of the propagator of SDEs
and their expansion for weak-noise are well established in the literature.
To set the notation and to keep the presentation self-contained, we summarize
here the essential features, following the formulation presented in \cite{chaichian2001}.
Let us consider the one-dimensional Langevin equation
\begin{equation}\label{EqLangevin}
\dot{v}=-f(v)+\sqrt{D}\xi(t),
\end{equation}
where $f(v)$ is a smooth force and
the Gaussian white noise $\xi(t)$ obeys Eq.~(\ref{ba}).
We can write down the conditional probability $ p(v_t,t|v_0,0) $ by using the path integral
formula \cite{chaichian2001},
\begin{equation}\label{EqPathIntegral}
p(v_t,t|v_0,0)=\int_{(v_0,0)}^{(v_t,t)} \mathcal{D}[v]  J[v]\mbox{e}^{-\frac{1}{4D}\int_0^t(\dot{v}+f(v))^2\mathrm{d}s} 
\end{equation}
where $ \int \mathcal{D}[v] $ denotes the Wiener measure
and the Jacobian term
\begin{equation}
J[v]=\exp\left( \frac{1}{2}\int_{0}^{t} f'(v) \mbox{d}s \right)
\label{eqappjac1}
\end{equation}
originates from the transformation $ \xi(t) \rightarrow v(t)$. Putting the two exponentials together, we thus express the kernel of the path integral in terms of the action
\begin{equation}\label{Eqaction}
S[v] = \int_{0}^{t} L(v(s),\dot{v}(s)) \mbox{d}s
=
\int_{0}^{t}
 \left((\dot{v}(s)+f(v(s)))^2-2Df'(v(s))\right) \mbox{d}s.
\end{equation}

All trajectories contribute to the path integral (\ref{EqPathIntegral}), but for small $ D $,
the largest contribution will come from the trajectory with smallest action.
At lowest order in $D$, this contribution is found by minimizing the action
\begin{equation}\label{EqzeroOrderAction}
S^{(0)}[v]=\int_{0}^{t} \left(\dot{v}(s)+f(v(s))\right)^2 \mbox{d}s ,
\end{equation}
which does not take the contribution of Jacobian into account
since it is multiplied by $D$. The corresponding
boundary value problem determined by the Euler-Lagrange equation reads
\begin{equation}\label{EqzeroOrderEL}
\ddot{v}^{(0)}(s) = f(v^{(0)}(s))f'(v^{(0)}(s)), \quad
v^{(0)}(0)=v_0, \quad v^{(0)}(t)=v_t.
\end{equation}
Given the path minimizing the action (\ref{EqzeroOrderAction}), the
leading order approximation of the propagator is thus given by
\begin{equation}\label{EqzeroOrderPropagator}
p^{(0)}(v_t,t|v_0,0)=N_1 \exp(-S^{(0)}[v^{(0)}]/(4D)),
\end{equation}
where the (time dependent) normalization $N_1$ can be computed a posteriori.

To improve this approximation, one can construct asymptotic
series expansions of the action, which do not, however, always ensure positivity of the
propagator. A simpler way to improve on the approximation (\ref{EqzeroOrderPropagator}) is to evaluate the action of
Eq.~(\ref{Eqaction}) with the Jacobian using the optimal path $ v^{(0)} $; see Eq.~(\ref{EqzeroOrderEL}). This leads to the expression
\begin{equation}\label{EqfirstOrderPropagatorZeroOrderPath}
p^{(0;1)}(v_t,t|v_0,0)=N_2 \exp(-S[v^{(0)}]/(4D)). 
\end{equation}
A more coherent approach, which keeps the spirit of the saddle-point approximation and
ensures positivity of the propagator, is to evaluate the minimizing path with the Jacobian, leading to the following Euler-Lagrange equation:
\begin{equation}\label{EqfirstOrderEL}
\ddot{v}^{(1)}(s) = f(v^{(1)}(s))f'(v^{(1)}(s))-D f''(v^{(1)}(s)), \quad
v^{(1)}(0)=v_0, \quad v^{(1)}(t)=v_t .
\end{equation}
In this case, the corresponding first order expression for the propagator is given by
\begin{equation}\label{EqfirstOrderPropagatorFirstOrderPath}
p^{(1)}(v_t,t|v_0,0)=N_3 \exp(-S[v^{(1)}]/(4D)) .
\end{equation}

In general, it is not possible to solve the boundary value problem of the
Euler-Lagrange equations (\ref{EqzeroOrderEL}) or
(\ref{EqfirstOrderEL}) analytically. Hence, numerical methods such as the shooting method
must be used. In this case, it is useful to respect the underlying Hamiltonian structure of the Euler-Lagrange problem by
using symplectic integration methods which preserve the Hamiltonian,
\begin{equation}\label{EqzeroOrderHamiltonian}
H^{(0)}(v,p_v)=p_v^2/2 - f^2(v)/2
\end{equation}
without the Jacobian or
\begin{equation}\label{EqfirstOrderHamiltonian}
H(v,p_v)=p_v^2/2- (f^2(v)+Df'(v))/2
\end{equation}
with the Jacobian. In this case, a symplectic Euler scheme  or the St\"{o}rmer-Verlet scheme, for example, can be
applied to integrate the corresponding canonical equations of motion.

\section{Energy-time relation}

\subsection{Monotonicity of $\theta_-$}
\label{AppMonotonicOfTheta}

Consider the expression defined in Eq.~(\ref{db}) either for $u_\tau<0<u_0$ and $H>0$ or for
$0<u_\tau<u_0$ and $H>H_{min}=-\tanh^2(u_\tau)/2$. The argument of the logarithm
in Eq.~(\ref{db}) can be written as
\begin{equation}\label{fa}
g(x)=\frac{a+\sqrt{x+a^2}}{b+\sqrt{x+b^2}}
\end{equation}
and it is easy to see that $g(x)>1$ in the given parameter ranges, i.e, either $b<0<a$ and $x>0$,
or $0<b<a$ and $x>-b^2$. In addition, it follows by differentiation
that $g'(x)<0$, i.e., $g$ is monotonically decreasing. Hence $\theta_-$
considered as a function of $H$ is the product of two positive
monotonically decreasing functions so that $\theta_-$ itself is monotonically decreasing.

\subsection{Critical points of $\theta_+$}\label{AppThreepath}

Since Eq.~(\ref{dd}) is a symmetric expression in the first two arguments it is sufficient to
consider the case $0<u_\tau<u_0$ and $H_{min}=-\tanh^2(u_\tau)/2<H<0$. The critical points are
determined by the vanishing derivative of $\theta_+$. It is more convenient to consider the
expression in terms of the new variable $\chi \in [0,u_\tau]$ defined by $H=-\tanh^2(\chi)/2$, where $-\chi$ represents the turning point of the indirect path,
introduced in Sec. \ref{case2}. Then differentiation gives
\begin{eqnarray}\label{EqtimePath3}
\frac{\partial \theta_+(u_0,u_\tau,-\tanh^2(\chi)/2)}{\partial \chi}
 =
h(\sinh(u_0),\sinh(\chi))+h(\sinh(u_\tau),\sinh(\chi)),
\end{eqnarray}
where we have introduced
\begin{equation}\label{fb}
h(a,z) = z\ln\left(a/z+\sqrt{(a/z)^2-1}\right) - \frac{(1+z^2)a}{z\sqrt{a^2-z^2}}, \quad
(0<z<a) .
\end{equation}
The critical points of $\theta_+$ are thus determined by the solutions of the equation
\begin{equation}
h(\sinh(u_0),z)+h(\sinh(u_\tau),z)=0.
\label{hsum1}
\end{equation} 
We now show that the right-hand side of Eq.~(\ref{EqtimePath3}) is a convex function of $z=\sinh(\chi)$, so there exist at most two solutions. To do so, compute the second derivative of Eq.~(\ref{fb}):
\begin{equation}\label{fc}
\frac{\partial h(a,z)}{\partial z} = \ln\left(a/z+\sqrt{(a/z)^2-1}\right)
+a \frac{z^4-2(a^2+1)z^2+a^2}{z^2 \left(a^2-z^2\right)^{3/2}}
\end{equation}
and
\begin{equation}\label{fd}
\frac{\partial^2 h(a,z)}{\partial z^2}=
-a \frac{z^4 (6+2 a^2)+z^2 (-5a^2+a^4)+2 a^4}{z^3 \left(a^2-z^2\right)^{5/2}}.
\end{equation}
It is easy to see that the numerator in Eq.~(\ref{fd}) is positive by completing the
square $6(z^2-a^2/2)^2$. Hence, the second derivative is negative for any (positive) values
of $a$. The same holds for Eq.~(\ref{hsum1}), proving convexity.

Finally, it is obvious from Eq.~(\ref{dd}) that $\theta_+(u_0,u_\tau,H)$ tends to $\infty$ as
$H\uparrow 0$ and that $\theta_+(u_0,u_\tau,H_{min})$ is finite and positive, resulting altogether
in a cubic shape for the graph of $\theta_+(u_0,u_\tau,H)$. 


\end{document}